\documentclass[12pt]{article}

\renewcommand{\theequation}{\arabic{section}.\arabic{equation}}

\def\hybrid{\topmargin -20pt    \oddsidemargin 0pt
        \headheight 0pt \headsep 0pt
        \textwidth 6.25in       
        \textheight 9.5in       
        \marginparwidth .875in
        \parskip 5pt plus 1pt   \jot = 1.5ex}

\hybrid

\catcode`\@=11

\def\marginnote#1{}
\newcount\hour
\newcount\minute
\newtoks\amorpm
\hour=\time\divide\hour by60
\minute=\time{\multiply\hour by60 \global\advance\minute by-\hour}
\edef\standardtime{{\ifnum\hour<12 \global\amorpm={am}%
        \else\global\amorpm={pm}\advance\hour by-12 \fi
        \ifnum\hour=0 \hour=12 \fi
        \number\hour:\ifnum\minute<10 0\fi\number\minute\the\amorpm}}
\edef\militarytime{\number\hour:\ifnum\minute<10 0\fi\number\minute}

\def\draftlabel#1{{\@bsphack\if@filesw {\let\thepage\relax
   \xdef\@gtempa{\write\@auxout{\string
      \newlabel{#1}{{\@currentlabel}{\thepage}}}}}\@gtempa
   \if@nobreak \ifvmode\nobreak\fi\fi\fi\@esphack}
        \gdef\@eqnlabel{#1}}
\def\@eqnlabel{}
\def\@vacuum{}
\def\draftmarginnote#1{\marginpar{\raggedright\scriptsize\tt#1}}

\def\draft{\oddsidemargin -.5truein
        \def\@oddfoot{\sl preliminary draft \hfil
        \rm\thepage\hfil\sl\today\quad\militarytime}
        \let\@evenfoot\@oddfoot \overfullrule 3pt
        \let\label=\draftlabel
        \let\marginnote=\draftmarginnote
   \def\@eqnnum{(\theequation)\rlap{\kern\marginparsep\tt\@eqnlabel}%
\global\let\@eqnlabel\@vacuum}  }

\def\preprint{\twocolumn\sloppy\flushbottom\parindent 2em
        \leftmargini 2em\leftmarginv .5em\leftmarginvi .5em
        \oddsidemargin -.5in    \evensidemargin -.5in
        \columnsep .4in \footheight 0pt
        \textwidth 10.in        \topmargin  -.4in
        \headheight 12pt \topskip .4in
        \textheight 6.9in \footskip 0pt
        \def\@oddhead{\thepage\hfil\addtocounter{page}{1}\thepage}
        \let\@evenhead\@oddhead \def\@oddfoot{} \def\@evenfoot{} }

\def\numberbysection{\@addtoreset{equation}{section}
        \def\theequation{\thesection.\arabic{equation}}}

\def\underline#1{\relax\ifmmode\@@underline#1\else
        $\@@underline{\hbox{#1}}$\relax\fi}

\def\titlepage{\@restonecolfalse\if@twocolumn\@restonecoltrue\onecolumn
     \else \newpage \fi \thispagestyle{empty}\c@page\z@
        \def\thefootnote{\fnsymbol{footnote}} }

\def\endtitlepage{\if@restonecol\twocolumn \else \newpage \fi
        \def\thefootnote{\arabic{footnote}}
        \setcounter{footnote}{0}}  

\catcode`@=12
\relax

\makeatletter
\newcounter{pubctr}
\def\publist{\@ifnextchar[{\@publist}{\@@publist}}
\def\@publist[#1]{\list
        {[\arabic{pubctr}]\hfill}{\settowidth\labelwidth{[999]}
        \leftmargin\labelwidth
        \advance\leftmargin\labelsep
        \@nmbrlisttrue\def\@listctr{pubctr}
        \setcounter{pubctr}{#1}\addtocounter{pubctr}{-1}}}
\def\@@publist{\list
        {[\arabic{pubctr}]\hfill}{\settowidth\labelwidth{[999]}
        \leftmargin\labelwidth
        \advance\leftmargin\labelsep
        \@nmbrlisttrue\def\@listctr{pubctr}}}
 \relax
\makeatother
\newskip\humongous \humongous=0pt plus 1000pt minus 1000pt

\newif\ifdtup

\relax

\def\d{\partial}

\def\sqr#1#2{{\vcenter{\vbox{\hrule height.#2pt\hbox{\vrule width.#2pt 
height#1pt \kern#1pt \vrule width.#2pt}\hrule height.#2pt}}}}

\def\=d{\,{\buildrel\rm def\over =}\,}

\def\i3p{\p32\int d^3p}

\def\As{A\hbox to 1pt{\hss /}}
\def\np4{\int d^4p_1\cdots d^4p_{n-1}\, }

\def\sgn{{\rm sgn}\, }

\def\P{\,{\rm P}}

\def\nx4{\int d^4x_1\ldots d^4x_n\, }

\def\kon#1#2{\vbox{\halign{##&&##\cr
\lower4pt\hbox{$\scriptscriptstyle\vert$}\hrulefill &
\hrulefill\lower4pt\hbox{$\scriptscriptstyle\vert$}\cr $#1$&
$#2$\cr}}}

\def\konv#1#2#3{\hbox{\vrule height12pt depth-1pt}
\vbox{\hrule height12pt width#1cm depth-11.6pt}
\hbox{\vrule height6.5pt depth-0.5pt}
\vbox{\hrule height11pt width#2cm depth-10.6pt\kern5pt
      \hrule height6.5pt width#2cm depth-6.1pt}
\hbox{\vrule height12pt depth-1pt}
\vbox{\hrule height6.5pt width#3cm depth-6.1pt}
\hbox{\vrule height6.5pt depth-0.5pt}}
\def\konu#1#2#3{\hbox{\vrule height12pt depth-1pt}
\vbox{\hrule height1pt width#1cm depth-0.6pt}
\hbox{\vrule height12pt depth-6.5pt}
\vbox{\hrule height6pt width#2cm depth-5.6pt\kern5pt
      \hrule height1pt width#2cm depth-0.6pt}
\hbox{\vrule height12pt depth-6.5pt}
\vbox{\hrule height1pt width#3cm depth-0.6pt}
\hbox{\vrule height12pt depth-1pt}}

\def\konw#1#2#3{\hbox{\vrule height12pt depth-1pt}
\vbox{\hrule height12pt width#1cm depth-11.6pt}
\hbox{\vrule height6.5pt depth-0.5pt}
\vbox{\hrule height12pt width#2cm depth-11.6pt \kern5pt
      \hrule height6.5pt width#2cm depth-6.1pt}
\hbox{\vrule height6.5pt depth-0.5pt}
\vbox{\hrule height12pt width#3cm depth-11.6pt}
\hbox{\vrule height12pt depth-1pt}}

\def\i{{\rm int}}

\def\e{{\rm ext}}

\def\r{{\rm ret}}
\def\a{{\rm av}}

\def\m3{{\mu_1\mu_2\mu_3}}

\def\co{{\rm Com}}

\def\p{{(+)}}

\def\be{\begin{equation}}       \def\eq{\begin{equation}}
\def\ee{\end{equation}}         \def\eqe{\end{equation}}

\def\bea{\begin{eqnarray}}      \def\eqa{\begin{eqnarray}}
\def\ena{\end{eqnarray}}        \def\eea{\end{eqnarray}}
                                \def\eqae{\end{eqnarray}}

\def\ba{\begin{array}}
\def\ea{\end{array}}
\def\unit{1 \hskip-.3em \raise2pt\hbox{$ \scriptstyle |$ } }

\def\a{\alpha}
\def\b{\beta}
 
\def\d{\delta}
\def\e{\epsilon}           
\def\f{\phi}               
\def\g{\gamma}
\def\h{\eta}   
\def\i{\iota}
\def\j{\psi}

\def\m{\mu}
\def\n{\nu}
\def\o{\omega}  
\def\p{\pi}                
\def\r{\rho}                                     
\def\s{\sigma}                                   
\def\t{\tau}

\def\D{\Delta}

\def\L{\Lambda}
  
\def\P{\Pi}

\def\cd{{\cal D}}

\def\cf{{\cal F}}

\def\cj{{\cal J}}
\def\ck{{\cal K}}
\def\cl{{\cal L}}

\def\co{{\cal O}}

\def\car{{\cal R}}

\def\ct{{\cal T}}

\def\half{{1 \over 2}}

\def\bop#1{\setbox0=\hbox{$#1M$}\mkern1.5mu
        \vbox{\hrule height0pt depth.04\ht0
        \hbox{\vrule width.04\ht0 height.9\ht0 \kern.9\ht0
        \vrule width.04\ht0}\hrule height.04\ht0}\mkern1.5mu}
\def\Box{{\mathpalette\bop{}}}                        
\def\pa{\partial}                              

\def\>{\rangle} 

\def\<{\langle} 
\def\Dsl{D \hskip-.6em \raise1pt\hbox{$ / $ } }

\def\sl#1{\rlap{\hbox{$\mskip 1 mu /$}}#1}
\def\leftrightarrowfill{$\mathsurround=0pt \mathord\leftarrow \mkern-6mu
       \cleaders\hbox{$\mkern-2mu \mathord- \mkern-2mu$}\hfill
       \mkern-6mu \mathord\rightarrow$}
\def\dvec#1{\vbox{\ialign{##\crcr
       \leftrightarrowfill\crcr\noalign{\kern-1pt\nointerlineskip}
       $\hfil\displaystyle{#1}\hfil$\crcr}}}          
\def\hook#1{{\vrule height#1pt width0.4pt depth0pt}}
\def\leftrighthookfill#1{$\mathsurround=0pt \mathord\hook#1
       \hrulefill\mathord\hook#1$}
\def\underhook#1{\vtop{\ialign{##\crcr                 
       $\hfil\displaystyle{#1}\hfil$\crcr
       \noalign{\kern-1pt\nointerlineskip\vskip2pt}
       \leftrighthookfill5\crcr}}}
\def\smallunderhook#1{\vtop{\ialign{##\crcr      
       $\hfil\scriptstyle{#1}\hfil$\crcr
       \noalign{\kern-1pt\nointerlineskip\vskip2pt}
       \leftrighthookfill3\crcr}}}

\def\sfrac#1#2{{\vphantom1\smash{\lower.5ex\hbox{\small$#1$}}\over
       \vphantom1\smash{\raise.4ex\hbox{\small$#2$}}}} 
\def\bfrac#1#2{{\vphantom1\smash{\lower.5ex\hbox{$#1$}}\over
       \vphantom1\smash{\raise.3ex\hbox{$#2$}}}}      
\def\afrac#1#2{{\vphantom1\smash{\lower.5ex\hbox{$#1$}}\over#2}}  
\def\on#1#2{{\buildrel{\mkern2.5mu#1\mkern-2.5mu}\over{#2}}}
\def\ddt#1{\on{\hbox{\LARGE .\kern-2pt.}}#1}             
\def\tdt#1{\on{\hbox{\LARGE .\kern-2pt.\kern-2pt.}}#1}   

\def\boxes#1{
       \newcount\num
       \num=1
       \newdimen\downsy
       \downsy=-1.5ex
       \mskip-2.8mu
       \bo
       \loop
       \ifnum\num<#1
       \llap{\raise\num\downsy\hbox{$\bo$}}
       \advance\num by1
       \repeat}
\def\boxup#1#2{\newcount\numup
       \numup=#1
       \advance\numup by-1
       \newdimen\upsy
       \upsy=.75ex
       \mskip2.8mu
       \raise\numup\upsy\hbox{$#2$}}

\newskip\humongous \humongous=0pt plus 1000pt minus 1000pt

\newif\ifdtup

\def\to{\rightarrow}

\def\1ov4{{1\over 4}}

\def\pa{\partial}
\def\xx{\times}

\def\dda{\dot{\alpha}} 
\def\ddb{\dot{\beta}}

\def\ddt{\dot{\t}}

\def\pa{\partial}

\def\xx{\times}

\def\nonu{\nonumber \\{}}
\def\half{{1 \over 2}}

\def\bfx{{\bf x}}
\def\bfy{{\bf y}}

\begin{document}


\thispagestyle{empty} \setcounter{footnote}{0}
\begin{flushright}
SLAC-PUB-7611 \\
ITP-SB-97-38\\
MPI/PhT-97-74\\
KUL-TF-97-18\\ 
hep-th/9803030
\end{flushright}

\vskip 1,7cm
\begin{center}
  \centerline{\large\bf Quantum Noether Method}

  \vskip 1.0cm 
{\bf Tobias Hurth \footnote{hurth@mppmu.mpg.de}}
  \vskip 0,2cm
  {\it Stanford Linear Accelerator Center, Stanford University\\
    Stanford, California 94309, USA}\\
  {\it and}\\
  {\it Max-Planck-Institute for Physics, Werner-Heisenberg-Institute\\
    F\"ohringer Ring 6, D-80805 Munich, Germany}
  \vskip 0,2cm
  {\bf Kostas Skenderis \footnote{kostas.skenderis@fys.kuleuven.ac.be    }}
  \vskip 0,2cm
  {\it Instituut voor Theoretische Fysica, KU Leuven\\
    Celestijnenlaan 200D, B-3001 Leuven, Belgium}

\end{center}
\vskip 0,5cm {\bf Abstract.} - {\small 
We present a general method for constructing perturbative quantum field
theories with global symmetries. We start from a free non-interacting 
quantum field theory with given global symmetries and we determine all 
perturbative quantum deformations assuming the construction
is not obstructed by anomalies. The method is established within the 
causal Bogoliubov-Shirkov-Epstein-Glaser approach to perturbative 
quantum field theory
(which leads directly to a finite perturbative series and does not 
rely on an intermediate regularization). Our construction 
can be regarded as a direct implementation of Noether's method 
at the quantum level. 
We illustrate the method by constructing the pure 
Yang-Mills  theory (where 
the relevant global symmetry is BRST symmetry), and the $N=1$ 
supersymmetric model of Wess and Zumino.
The whole construction is done before the so-called adiabatic 
limit is taken. Thus, all considerations regarding symmetry, unitarity 
and anomalies are well-defined even for massless theories.}
\vskip 0,2cm

\newpage

\section{Introduction}
\setcounter{equation}{0}

Symmetries have always played an important and fundamental r\^{o}le
in our quest of understanding nature. In classical physics 
they lead to conserved quantities (integrals of motion).
The latter constrain the classical evolution of the system 
and some times they even uniquely determine it (for example, 
in the case of two-dimensional exactly solvable models). 
The r\^{o}le of symmetries in quantum physics is equally important. 
In quantum field theory symmetries lead to relations among the Green 
functions of the theory (Ward-Takahashi identities). The latter are  
instrumental in the proof of renormalizability and unitarity of the theory 
under question. Furthermore, recent developments in supersymmetric 
theories\cite{SW} show that global symmetries themselves are sometimes 
sufficient to determine the structure of the theory. 
Thus, it seems  desirable to carefully 
understand the inter-relations between symmetries and quantum theory in a 
manner which is free of the technicalities inherent in the conventional 
Lagrangian approach (regularization-renormalization), and also in a way 
which is  model independent as much as possible. In this article we shall
undertake a first step towards this goal. We shall analyze this question
within perturbative quantum field theory(QFT). 

One may argue that, to a large extent, the relation between symmetries and 
perturbative QFT is by now well-understood. However, one would like 
to have an understanding at a more fundamental level. Namely, to separate
the generic properties that symmetries impose from the specifics
of a given model that realizes this symmetry. In addition, it would desirable
to have a formulation which is mathematically as sound as possible. 
 
A framework that encompasses most of the desired properties for this 
kind of questions is the causal 
approach to perturbative quantum field theory introduced by 
Bogoliubov and Shirkov \cite{BS} and developed by 
Epstein and Glaser\cite{EG0,EG,EG2,stora}.
The explicit construction method of Epstein and Glaser rests
directly on the axioms of relativistic quantum field theory. 
On the one hand, it clarifies how the
fundamental axioms guide the perturbative construction of the $S$
matrix, and how well-defined time-ordered products are directly
constructed
without the need of an intermediate regularization of the theory. 
On the other hand, it is an explicit construction method
for the most general perturbation series compatible with causality and
Poincar\'{e} invariance.  The purely technical details which are 
essential for explicit calculations are separated from the 
simple physical
structure of the theory. With the help of the causality condition, the 
well-known problem of ultraviolet (UV)
divergences is reduced to a mathematically well-defined problem,
namely the splitting of an operator-valued distribution with causal
support into a distribution with retarded and a distribution with
advanced support or, alternatively \cite{stora, fredenhagen1}, 
to the continuation of time-ordered
products to coincident points. 
Implicitly, every consistent renormalization scheme
solves this problem. In this sense the explicit Epstein-Glaser 
(EG) construction should not be regarded as a special renormalization 
scheme but as a general framework in which the conditions posed by the 
fundamental axioms of QFT on any renormalization scheme are built in by
construction. In the EG approach the $S$-matrix is directly
constructed in the well-defined Fock space of free asymptotic fields
in the form of a formal power series. Thus, one does not need the
Haag-Ruelle \mbox{(LSZ-)} formalism. Interacting field operators can still
be perturbatively constructed in an additional step as certain
functional derivatives of the $S$-matrix (\cite{EG} section 8, \cite{BKS}, 
see also appendix B).  
 
In classical physics, Noether's theorem states that there is a
conserved current for every invariance of the classical action under a
continuous, internal or spacetime symmetry transformation of the
fields. This theorem also allows for  an iterative method to construct
invariant actions, called the Noether method 
\cite{deser}. Noether's method has been
used in the construction of theories with local symmetries starting from 
ones with only rigid symmetries. For example, this method 
was extensively used in the construction of
supergravity theories \cite{sugra}.
In a slightly different setting, starting from a free Lagrangian one
can iteratively construct interactions
by adding extra terms to the action and to
transformation rules in such a way that 
the final action is invariant under the 
modified transformations.
One may try to elevate these results to the quantum regime by 
quantizing the system. To this end, one should investigate 
the compatibility of the classical symmetry with the quantization.
The latter is reflected in the absence or presence of anomalies. 

In this paper we propose a general quantum method which, as we shall
see, is a direct implementation of Noether's method at the quantum
level. For this reason we shall call it ``Quantum Noether Method''.
Starting from a free quantum field theory, well-defined 
in the Fock space of free asymptotic fields, the method allows for 
a construction of all perturbative quantum theories.
Such a direct implementation of the Noether method
in the quantum theory is established in the causal Epstein-Glaser 
approach to perturbative quantum field theory. 
In the Quantum Noether Method  the conditions for constructing
a classical action and the conditions for 
absence of anomalies are associated with obstructions in the 
construction of the $S$-matrix. The classical action emerges from 
the cancellation of tree-level obstructions, whereas anomalies are 
associated with loop  obstructions. An algebraic consistency 
condition for possible obstructions can be derived without using the quantum
action principle\cite{paper2}.  

The strength of the EG construction lies in the operator
formalism.  The proof of general properties of a given quantum
field theory can be simply and also rigorously reduced to the discussion
of the local normalization ambiguity which is restricted by power
counting. It is also the operator formalism which circumvents
the classical problem of overlapping divergences - in the usual framework
the latter problem is solved by the famous forest formula.
Moreover, the EG formalism provides a natural framework to discuss 
symmetries in perturbative quantum field theories in which the 
regularization and scheme independence of anomalies as well as 
the reduction of their discussion to local normalization 
ambiguities is manifest. From this it is clear that the Epstein-Glaser 
approach  provides an ideal framework 
for the discussion of symmetries in quantum field theory.  

In the original article \cite{EG} Epstein and Glaser applied their
construction to scalar field theory.  The extension of these results
to abelian  gauge theory in the four dimensional Minkowski space
has been worked out some years ago in \cite{Blanchard,dos,DHS93,Scharf95}. 
The causal Epstein-Glaser construction of
(3+1)-dimensional non-abelian gauge theory in the Feynman gauge
coupled to fermionic matter fields was performed  in \cite{DHS95,H95}.
There, a definition of non abelian gauge invariance was given as an operator
condition in every order of perturbation theory separately.
This condition involved only the linear (abelian) BRST-transformations
of the free asymptotic field operators. 
It was claimed that this operator condition expresses the
whole content of non-abelian gauge structure in perturbation theory
\cite{DHS95,H95}. In fact, it was proven that the operator condition
directly implies the unitarity of the $S$-matrix in the physical
subspace, i.e.  decoupling of the unphysical degrees of freedom.
Furthermore, it was shown that from the operator condition one can
derive the Slavnov-Taylor identities for the connected
Green functions. 

Up until now, however, only the  Yang-Mills (YM) theory was fully
constructed as a quantum theory in this framework.  Moreover, a deep
understanding of how the asymptotic operator condition develops the
full BRST symmetry\cite{BRST}  was missing. 
With this ingredient missing, it was
not {\it a priori} clear whether the quantum theory constructed using
the EG procedure actually coincides with the usual YM theory or it is
some kind of ``Yang-Mills-like'' theory which cannot be reached from
the conventional Lagrangian approach. Since  the condition of
asymptotic BRST invariance seems to be  a weaker condition than
the full BRST symmetry it was not ruled out
that there are new theories compatible with this symmetry - an
interesting possibility because the asymptotic symmetry condition was
shown to be sufficient for decoupling of the unphysical degrees of
freedom. We shall argue,
however, that the full BRST transformations were already present in
the analysis of \cite{DHS95,H95}, thereby establishing that the theory
constructed by the EG procedure coincides with the usual YM theory.

Having understood fully the case of YM theories, we are in the
position to generalize the construction to any theory with any local
and global symmetry. Notice that local symmetries manifest themselves
through the rigid BRST symmetry. Thus, one can treat both cases in
parallel. Starting with the condition of asymptotic symmetry we show
explicitly how the formalism automatically develops the 
anomaly-free full
quantum symmetry of the interacting system, provided such an 
anomaly-free deformation of the given free theory exists.    
In this paper we exclude any 
possible obstructions of the symmetry at tree and loop level. 
A systematic cohomological analysis of possible  obstructions 
will be presented in a separate paper \cite{paper2}. The Epstein-Glaser
framework  allows for a simplified derivation of algebraic consistency
conditions for obstructions. Moreover, the cohomological analysis
of such obstructions is established before the adiabatic limit   
is taken. So the discussion is also applicable to massless
theories such as Yang-Mills theories \cite{paper2}.

The idea behind the construction is very simple. Given a set of free
fields and a symmetry (such as supersymmetry or BRST symmetry)
generated by a Noether current (which at this point only captures
the linear part of the transformation rules), 
one demands that the current is conserved at the quantum level, 
i.e. inside correlation functions (Quantum Noether Condition).
We shall show that this condition at tree-level automatically produces 
the most general non-linear completion of the transformation rules 
and also the corresponding Lagrangian which is invariant under these 
transformation rules. We shall then examine the Quantum Noether Condition at 
loop level. We shall show that if the anomaly consistency condition
has only trivial solutions then the theory is stable, i.e. all local 
terms that are produced by loops are already present at tree-level.
This we shall call generalized renormalizability 
(this corresponds to the notion of ``renormalizability in the modern sense''
introduced for gauge theories  in \cite{GW95}).
If in addition the theory is also power counting renormalizable 
then generalized 
renormalizability coincides with the usual renormalizability.
The only restriction needed for our construction to work
is that the power counting index (singular order in EG, see section 3) 
is bounded in every order in perturbation theory. So, in particular,
our consideration also apply to effective field theories that are 
not power counting renormalizable. 

It is rather remarkable that the only information one needs in order
to construct a perturbative quantum field theory with a given 
global symmetry is a set of free fields linearly realizing this symmetry
(which is assumed to be generated by a Noether current, see 
footnote \ref{restriction}).
Even the first term in the $S$-matrix, which is usually regarded as
an input in the EG formalism is now derived using the Quantum Noether
Condition.
 
We have organized this paper as follows. In section 2 we shortly recall
the Noether method. In section 3  
we provide a self-contained summary of the basic ingredients of the 
EG construction.  Section 4 is the main section where we establish the 
Quantum Noether Method. We illustrate the method in section 5 with 
two examples; the case of pure  Yang-Mills theory and the $N=1$
supersymmetric Wess-Zumino model. Special care was taken  
in order to illustrate every step of the general 
construction explicitly. In Appendix A we explain our conventions 
in detail. In Appendix B  we discuss several issues such as 
the infrared problem, the construction of interacting fields, 
the problem of overlapping divergences that further motivate 
the use of the EG formalism.

\section{The Noether Method\\
 in Classical Field Theory}
\setcounter{equation}{0}

In this section we shortly recall the Noether method. 
Let us start with 
a classical Lagrangian density $\cl (\f^A, \pa_\m \f^A)$ 
that depends on a number of fields (both bosons and fermions) 
$\f^A$ and their first derivative $\pa_\m \f^A$, where $A$ is an index that
distinguishes different types of fields.  Suppose now that the
action $S=\int \cl$ is invariant  under the symmetry transformation $s \f^A$. 
This means that the Lagrangian density 
transforms into a total derivative, $s \cl = \pa_\m k^\m$.
A standard way to derive Noether's current is to let the parameter of
the symmetry transformation $\e$ become local. Then the Noether
current is the expression multiplying the derivative of the local
parameter.  
\be \label{nothdef}
\d S = \int j^\m (\pa_\m \e)  
\ee 
Taking the parameter $\e$ rigid one sees that the variation is 
indeed a symmetry of the
action. On the other hand, if the field equation are satisfied, 
$\d S=0$ for any $\e$ and, therefore,
$\pa_\m j^\m=0$.  

The Noether current is given by 
\be 
\label{Noe2}
j^\m = \frac
{\pa \cl}{\pa (\pa_\m  \f^A)} s \f^A - k^\m. 
\ee 
We shall always include the parameter of the transformation in the 
current. In this way the current is always bosonic.
Direct calculation (using $\pa_\m k^\m = s \cl$) yields,
\be \label{jdiv}
\pa_\m j^\m = [\pa_\m {\pa \cl \over \pa(\pa_\m \f^A)} - 
{\pa \cl \over \pa \f^A}] s \f^A.
\ee
Clearly, $\pa_\m j^\m=0$ when the field equations are satisfied.
There is a natural arbitrariness in the definition of the current.
One may always add terms of the form $\pa_\m b^{\mu \nu}$, 
where $b^{\mu \nu}$ is antisymmetric in $\mu,\nu$. 

The conserved charge $Q$ is equal to 
\be 
Q = \int d^3x (p_A s \f^A - k^0), 
\ee 
where $p_A=\pa \cl/\pa_0 \f^A$ is the conjugate momentum of
$\f^A$.  One may check that $Q$ generates the
corresponding variation when acting (by the Poisson bracket) to the
fields,  
\be \label{Qpois}
\{ Q, \f^A \} = s \f^A.  
\ee 

As mentioned previously, Noether's theorem allows for an
iterative method to construct invariant actions, called the Noether
method \cite{deser}. Starting from a free Lagrangian one can iteratively
construct interactions in classical field theory by adding extra terms
to the action and to the transformation rules such that the final action
is invariant.  The way the Noether method works is as follows. 
Start from an action $S_0 = \int \cl_0$ invariant under transformations
$s_0 \f^A$. This set is assumed to be closed on-shell.
The goal is then to find a new action, 
\be 
S=\int d^4x (\cl_0 + g \cl_1 + g^2 \cl_2 + \cdots), 
\ee and new transformation rules, 
\be  
s \f^A = s_0 \f^A + g s_1 \f^A + g^2 s_2 \f^A + \cdots, \label{tr1} 
\ee 
where $g$ is a new coupling constant (or deformation
parameter), such that the new action is invariant under the new
transformation rules.  To first order in $g$ the relevant equation
reads 
\be 
\frac{\d \cl_0}{\d \f^A} (s_1 \f^A) + \frac{\d \cl_1}{\d \f^A} 
(s_0 \f^A) -\pa_\m k^\m_1 =0 \label{step1} 
\ee 
Note that this is an equation for $s_1, \cl_1$ and $k^\mu_1$ which
may be not solvable (which means there is no possible deformation). 
Starting with an ansatz for $s_1 \f^A$ one 
tries to determine a $\cl_1$ such that the above 
equation holds or vice versa.  If a solution $(s_1,\cl_1,k^\mu_1)$ of 
equation (\ref{step1}) is found,  one tries to solve the equation that 
appears at order $g^2$, and so on. The corresponding current 
is given by 
\bea
j^\m &=& 
[(\frac{\pa \cl_0}{\pa (\pa_\m \f^A)} s_0 \f^A  -k_0^\m)  
+ (g \frac{\pa \cl_1}{\pa (\pa_\m \f^A)} 
+ g^2 \frac{\pa \cl_2}{\pa (\pa_\m \f^A)} + \cdots) s_0 \f^A] \nonu
&&+ g [(\frac{\pa \cl_0}{\pa (\pa_\m \f^A)} s_1 \f^A  - k_1^\m)
+ (g \frac{\pa \cl_1}{\pa (\pa_\m \f^A)} +\cdots) s_1 \f^A] 
\cdots \label{noether}
\eea
where we have organized the terms in a way that it will be useful 
in later sections.

A systematic way to organize this procedure is to use the 
Batalin-Vilkovisky (or anti-field) formalism\cite{BV}. 
Although we will not use this formulation in the present article
we provide a short description of the method as it provides 
a nice reformulation of the problem\footnote{In this article we 
do not discuss 
the precise conditions under which  the classical Noether method and the 
reformulation using anti-fields are equivalent.}. 
We refer to the literature \cite{antiN} for a more detailed description. 

In the anti-field formalism (for a detailed
exposition see \cite{HT}) one first replaces
the parameter of the symmetry variation by a ghost field
(in the case of global symmetries the latter is a constant field \cite{BHW})
and introduces a new field, the anti-field $\f_A^*$, for each field $\f^A$
(the fields $\phi^A$ include the ghost field).  
The anti-fields act as sources for symmetry variation of the corresponding
field, namely one adds in the Lagrangian a term $\f^*_A s \f^A$
(but the solution of (\ref{master}) may contain higher powers of anti-fields). 
The defining equation of the theory is the master equation 
\be \label{master}
(S, S)=0,
\ee 
where $(A, B)$ denotes the anti-bracket, 
\be (A, B) = {\stackrel{\leftarrow}{\pa}A \over \pa \f^A} 
\frac{\stackrel{\rightarrow}{\pa}B}{\pa \f_A^*} 
-{\stackrel{\leftarrow}{\pa}A \over \pa \f_A^*} 
\frac{\stackrel{\rightarrow}{\pa}B}{\pa \f^A},  
\ee 
where the arrow indicates from where the derivative acts.
The action in this formalism generates the transformation
of the fields, 
\be 
s \f^A = (\f^A ,S). \label{tr2}
\ee 

The problem is now formulated as follows\cite{antiN}. 
Starting from an $S_0$ that solves the master equation $(S_0, S_0)=0$
(for a theory with a closed gauge algebra
$S_0 = \int (\cl_0 + \f^*_A s_0 \f^A$)) 
one seeks for a new action $S = S_0 + g S_1 + g^2 S_2 + \cdots$,
where $S_1, S_2,...$ are local functionals,
that solve the new master equation $(S, S)=0$. This equation
yields a tower of equations once it is expanded in $g$. The first few
equations are the following, 
\bea
&&(S_0, S_1) =0, \label{bv1} \\
&&2 (S_0, S_2) + (S_1, S_1) =0, \label{bv2} 
\eea 
and so on. Solving these equations  
one obtains both the new terms in the action and the
new transformation rules. The latter are obtained by using
(\ref{tr2}). Let us define the derivations
\be \label{deriv} 
s_i =(\ \ , S_i).  
\ee 
These derivations correspond to the ones in (\ref{tr1}). 
Notice, however, that in this formulation there may be constant ghosts 
present, so one might have to first eliminate them in order to compare
(\ref{deriv}) and (\ref{tr1}).

A nice feature of this approach is that it allows for a systematic
cohomological approach to the problem. For instance, equation (\ref{bv1}) 
tells us that $S_1$ is an element of $H^0(s_0 ,d)$, where $H^k(s_0,d)$
denotes the  cohomology group of the differential $s_0$ relative to
the differential $d$ in ghost number $k$ in the space of local
functionals. In addition, 
the obstruction to the solvability of this equation lies in
$H^1(s_0, d)$. Similar remarks apply for the rest of the equations.

\section{The Causal Method of Epstein-Glaser}
\setcounter{equation}{0}

We shall give a short but self-contained introduction to the 
causal Epstein-Glaser construction. 
This section may serve as a glossary for all quantities
we are about to use in the next section. 
For some of the technical details we refer  
to the literature \cite{EG,EG2,stora,Scharf95}.
We note that we follow the original Epstein-Glaser article \cite{EG} 
in our  presentation. So  we differ slightly from reference 
\cite{Scharf95,H95} regarding the causality condition,
and the role of the Wick submonomials in the construction.

\subsection{Inductive Construction}

We recall the basic steps of the Epstein-Glaser construction in the
case of a massless scalar field. For concreteness, we consider the 
case of four dimensional spacetime. The formalism, however, is valid
in any dimension. The very starting point is the Fock
space ${\cal F}$ of the massless scalar field (based on a
representation space $H_s^{m=0}$ of the Poincar\'{e} group)
with the defining equations
\begin{equation}
\Box \varphi  = 0  \quad {\bf (a) } ,\quad 
[\varphi (x),\varphi (y)] = i \hbar D_{m=0} (x-y) \quad  {\bf (b)},
\end{equation}
where $D_{m=0} (x-y)= \frac{-i}{(2\pi)^3} \int dk^4 \delta(k^2) 
\sgn(k^0) \exp(-ikx)$ is the zero-mass Pauli-Jordan distribution 
(see appendix A).
In contrast to the Lagrangian approach, the $S$-matrix is directly
constructed in this Fock space in the form of a formal power series
\begin{equation}
\label{GC}
 S(g) = 1 + \sum_{n=1}^\infty \frac{1}{n !} 
\int dx_1^4 \cdots dx_n^4 \quad 
T_n(x_1, \cdots, x_n; \hbar) \quad  g(x_1) \cdots g(x_n).
\end{equation}
(We do not include explicit $i$ factors in (\ref{GC}) in order 
to reduce the number of $i$-factors in our equations.)
In this approach the coupling constant $g$ is replaced by a 
tempered test function $g(x) \in {\cal S}$ (i.e. a smooth function
rapidly decreasing at infinity) which switches on the interaction. 

The central objects are the $n$-point operator-valued 
distributions\footnote{$T_n \in {\cal S'}$, where ${\cal S'}$
 denotes the space of functionals on ${\cal S}$.} $T_n$. 
They should be viewed as mathematically
well-defined (renormalized) time-ordered products,
\begin{equation}
 T_n(x_1, \cdots, x_n; \hbar) = T \left[T_1(x_1) \cdots T_1(x_n)\right],
\end{equation}
of a given specific coupling, say 
$T_1={i \over \hbar} :\Phi^4: \quad {\bf (c)},$ 
which is the third defining equation in order to
specify the theory in this formalism. 

Notice that the expansion in (\ref{GC}) is {\it not} a loop expansion.
Each $T_n$ in (\ref{GC}) can receive tree-graph and loop-contributions.
One can distinguish the various contributions from the power of 
$\hbar$ that multiplies them\footnote{
The fact that the distributions $T_n$ are formal Laurent series in $\hbar$ 
follows from the way $\hbar$ appears in defining equations (b) and (c) and the 
explicit construction that we describe below. 
Furthermore, one may deduce that $\hbar$ is a loop 
counting parameter (for connected graphs)
by using similar arguments as in the Lagrangian 
formulation. In particular, connected 
tree-level graphs come with a factor of 
$1/\hbar$, 1-loop graphs with $\hbar^0$, etc.}. 

Epstein and Glaser present an
explicit inductive construction of the most general perturbation series
in the sense of (\ref{GC}) which is compatible with the fundamental
axioms of relativistic quantum field theory, causality and 
Poincar\'{e}
invariance, which
can be stated as follows:\\
$\bullet$ Let $g_1$ and $g_2$ be two tempered test functions. Then
causal factorization means that
\begin{equation}
\label{GD}
S(g_1 + g_2) = S(g_2)  S(g_1)    \quad if \quad 
\mbox{supp} g_1 \preceq \mbox{supp} g_2
\end{equation}
the latter notion means that the support of $g_1$ and the support of
$g_2$, two closed subsets of ${\bf R}^4$, can be separated by a space
like surface; more precisely $\mbox{supp} g_2$ does not intersect the past
causal shadow of $\mbox{supp} g_1$: 
\begin{equation}
\mbox{supp} g_2 \cap ( \mbox{supp} g_1 + \bar{V}^-) = 0,
\end{equation}
\begin{equation}
(\bar{V})^- = \{ x \in {\bf R^4} | x^0 \leq |\vec{x}| \}
\end{equation}
$\bullet$ Let $U(a,{\bf \Lambda} )$ be the usual representation
of the Poincar\'{e} group $P_+^4$ in the given Fock space ${\cal F}$. Then
the condition of Poincar\'{e} invariance of the $S$-matrix says that
\begin{equation}
\label{Poi}
U(a,{\bf \Lambda} ) S(g) U(a,{\bf \Lambda} )^{-1} = 
S(g_\Lambda^a) \quad \forall a \in {\bf R}^4, 
\forall {\bf \Lambda} \in L_+^4, g_\Lambda^a (x) = g({\bf \Lambda}^{-1}(x-a))
\end{equation}
Actually, in order to establish the general construction 
only translational
invariance is needed. Lorentz invariance can be imposed in addition
in a subsequent
step.

It is  well-known that the heuristic solution for (\ref{GD}), namely
\begin{equation}
\label{heuristic}
T_n(x_1, \ldots, x_n; \hbar) = \sum_{\pi} T_1(x_{\pi (1)} ) \ldots 
T_1(x_{\pi (n)} )  \Theta (x_{\pi (1)}^0 - x_{\pi (2)}^0 ) \ldots 
\Theta (x_{\pi (n-1)}^0 - x_{\pi (n)}^0),
\end{equation}
is, in general, affected by 
ultra-violet divergences ($\pi$  runs over all permutations of 
$1, \ldots, n$).
The reason for this is that the
product of the discontinuous $\Theta$-step function with Wick monomials
like $T_1$ which are operator-valued distributions is ill-
defined. One
can handle this problem by using the usual regularization and
renormalization procedures and finally end up with the 
renormalized time-ordered products of the couplings $T_1$.

Epstein and Glaser suggest another path which leads directly to 
well-defined $T$-products without any intermediate modification of
the theory using the fundamental property of causality (\ref{GD}) as a
guide. They translate the condition (\ref{GD}) into an induction
hypothesis, $H_m, m<n$, for the $T_m$-distribution which reads 
\be \label{caus1}
H_m:\left\{
\ba{c}
\hspace{0.1cm} T_m(X \cup Y) = T_{m_1} (X) \ T_{m-m_1} (Y) \quad {\rm if}  
\quad X \succeq  Y,\quad X,Y \neq \emptyset, \ 0<m_1<m \\
\hspace{0.1cm} [T_{m_1} (X), T_{m_2} (Y)] = 0 \ {\rm if} \ X \sim Y \
(\Leftrightarrow X \succeq Y \wedge X \preceq  Y) \ \forall m_1,m_2 \le m
\ea
\right.
\ee
Here we use the short-hand notation $ T_m(x_1, \ldots, x_m; \hbar) = T(X)$;
$ \mid X \mid = m $.\\
Besides other properties they also include the Wick formula for the
$T_m$ distributions into the induction hypothesis. 
This is most easily done by including the so-called Wick submonomials  
of the specific coupling $T_1 =(i/\hbar) :\Phi^4:$ as
additional couplings in the construction 
$T_1^j := (i/\hbar) (4!/(4-j)!) :\Phi^{4-j}:, 0<j<4$.
Then the Wick formula for the $T_n$ products can be written as
\be
T_m [ T^{j_1}_1(x_1) \cdots T^{j_m}_1(x_m)]
=\!\!\!\!\sum_{s_1,..,s_m}\<0\mid T[ T^{j_1+s_1}_1(x_1)\cdots 
T^{j_m+s_m}_1(x_m)]   \mid 0\>
:\prod_{i=1}^m[\frac{\Phi^{s_i}(x_1)} {s_i !}]: 
\ee 
In short-hand notation the formula reads
\begin{equation}
\label{Wick}
T_m^{\vec{j}} (X) = \sum_{\vec{s}} 
\< 0 \mid T_m^{\vec{j}+\vec{s}} (X) \mid 0 \> \frac{:\Phi^{\vec{s}}: (X)}{\vec{s}}
\end{equation}
That such a quantity is a well-defined operator-valued distribution in
Fock space is assured by distribution theory (see Theorem O in
\cite{EG}, p. 229).  Note also that the coefficients in the Wick expansion are
now represented as
vacuum expectation values of operators. 

Now let us assume that $T_m$ distributions with all required
properties are successfully constructed for all $m<n$.
Epstein and Glaser introduce then the retarded and the
advanced $n$-point distributions (from now on, in this section, 
we suppress the $\hbar$ factor in our notation):
\begin{equation}
\label{ret}
R_n(x_1,\ldots ,x_n)=T_n(x_1,\ldots ,x_n)+R'_n, 
\quad R'_n=\sum_{P_2} T_{n-n_1}(Y,x_n)\tilde{T}_{n_1}(X) 
\end{equation}
\begin{equation}
\label{adva}
A_n(x_1,\ldots ,x_n)=T_n(x_1,\ldots ,x_n)+A'_n, 
\quad A'_n=\sum_{P_2} \tilde{T}_{n_1}(X)T_{n-n_1}(Y,x_n).
\end{equation}
The sum runs over all partitions $P_2:\{x_1,\ldots x_{n-1} \}=X \cup
Y, \quad X \not= \emptyset$ into disjoint subsets with $\mid X \mid
=n_1 \ge 1, \mid Y \mid \le n-2.$
The $\tilde{T}$ are the operator-valued distributions of the inverse S-matrix:
\begin{equation}
S(g)^{-1}=1+\sum_{n=1}^\infty \frac {1}{n!} \int d^4x_{1}\ldots 
d^4 x_n \tilde{T}_n(x_1,\ldots x_n)g(x_1)\ldots g(x_n)
\end{equation}
The distributions $\tilde{T}$\
can be computed by formal inversion of $S(g)$:
\begin{equation}
S(g)^{-1}=(\bf 1 \rm + T)^{-1}=\bf 1\rm + \sum_{n=1}^\infty (- T)^r
\end{equation}

\begin{equation}
\tilde{T}_n(X)=\sum_{r=1}^n (-)^{r} \sum_{P_r}T_{n_1}(X_1)
\ldots T_{n_r}(X_r),
\end{equation}
where the second sum runs over all partitions $P_r$ of X into r disjoint 
subsets $X=X_1\cup\ldots\cup X_r,\quad X_j\not=\emptyset,\quad 
\mid X_j \mid =n_j.$

We stress the fact that all products of distributions are well-defined 
because the arguments are disjoint sets of points so that 
the products are tensor products of distributions.
We also remark that both sums, $R'_n$ and $A'_n$,  in contrast to $T_n$, 
contain $T_j$'s with $j \le n-1$ only
and are therefore known quantities in the inductive step from $n-1$ to
$n$. Note that the last argument $x_n$ is marked as the reference point for
the support of $R_n$ and $A_n$.
The following crucial support property is a consequence of the
causality conditions (\ref{caus1}):
\begin{equation} \label{support1}
\mbox{supp} R_m(x_1,\ldots ,x_m) \subseteq \Gamma_{m-1}^+(x_m), \quad m < n 
\end{equation}
where $\Gamma_{m-1}^+$ is the ($m-1$)-dimensional closed forward cone,
\begin{equation}
\Gamma_{m-1}^+(x_m)=\{(x_1,\ldots ,x_{m-1}) 
\mid (x_j - x_m)^2 \ge 0, x_j^0 \ge x_m^0, \forall j \}.
\end{equation}
In the difference
\begin{equation}
\label{D-dist}
D_n(x_1, \ldots ,x_n) \=d  R'_n-A'_n
\end{equation}
the unknown $n$-point distribution $T_n$ cancels. Hence this quantity
is also known in the inductive step.  With the help of the causality
conditions (\ref{caus1}) again, one shows that $D_n$
has causal support
\begin{equation}
\mbox{supp} D_n \subseteq \Gamma_{n-1}^+(x_n) \cup \Gamma_{n-1}^-(x_n) 
\end{equation}
Thus, this crucial support property is preserved in the 
inductive step from $n-1$ to $n$. 

Given this fact, the following inductive construction of the $n$-point
distribution $T_n$ becomes possible: Starting off with the known
$T_m(x_1, \ldots , x_n)$, $m \le n-1,$ one computes $A'_n, R'_n$
and $D_n = R'_n - A'_n.$ With regard to the supports, one can
decompose $D_n$ in the following way:
\begin{equation}
D_n(x_1, \ldots , x_n) = R_n (x_1, \ldots , x_n) - 
A_n (x_1, \ldots , x_n)
\end{equation}
\begin{equation}
\mbox{supp} R_n \subseteq \Gamma_{n-1}^+(x_n), \quad \mbox{supp} A_n 
\subseteq \Gamma_{n-1}^-(x_n)
\end{equation}
Having obtained these quantities we define $T'_n$ as
\begin{equation}
\label{TTT}
T'_n = R_n - R'_n = A_n - A'_n
\end{equation}
Symmetrizing over the marked variable $x_n$, we finally obtain 
the desired $T_n$,
\begin{equation}
T_n(x_1,\ldots x_n)=\sum_{\pi} \frac{1}{n!} 
T'_n(x_{\pi (1)}, \ldots x_{\pi (n)}) 
\end{equation}
One can verify that the $T_n$ satisfy the conditions (\ref{caus1}) 
and all other further properties of the induction
hypothesis \cite{EG}.

\subsection{Distribution Splitting}

Le us now discuss the splitting the operator-valued distribution $D_n$.
As  follows from our discussion this is 
the only nontrivial step in the construction.

Let there be an operator-valued tempered 
distribution  $D_n \in  \cal S' ({\bf R^{4n}})$ 
with causal
support,
\begin{equation}
\mbox{supp} D_n \subseteq \Gamma_{n-1}^+(x_n) \cup \Gamma_{n-1}^-(x_n). 
\end{equation}
then the question is whether it is possible to find a pair (R, A) of
tempered distributions on ${\bf R^{4n}}$ with the following
characteristics:
\begin{equation}
\bullet\quad R, A \in  \cal S \it' (\bf R^{4n})\qquad \mbox{\bf (A)}
\end{equation}
\begin{equation}
\bullet\quad \mbox{supp} R \subset \Gamma^+(x_n), \quad \mbox{supp} A 
\subset \Gamma^-(x_n)\qquad \mbox{\bf (B)}
\end{equation}
\begin{equation}
\bullet\quad R - A = D\qquad \mbox{\bf (C)}
\end{equation}
The EG formalism reduces the usual renormalization program to this
mathematically well-defined problem. Every
renormalization scheme solves this problem implicitly. For example, 
the well-known BPHZ renormalization scheme which is often regarded 
as the most solidly founded explicit renormalization scheme also defines 
a splitting solution.
As mentioned already in the introduction, there is a 
complication, namely the well-known problem
of overlapping divergences which is solved by the
famous forest formula in the BPHZ framework. The EG formalism provides 
a natural solution to this problem 
by implementing the causality condition directly
on the operator level (see Appendix B).  

The problem of distribution splitting has been solved in a general
framework by the mathematician Malgrange in 1960 \cite{Malgrange}.
Epstein and Glaser used his general result for the special case of quantum 
field theory. A new solution  of the splitting problem was given 
recently \cite{Scharf95}.

We mention that the Wick formula for $T$-products, (\ref{Wick}) 
$\forall m<n$, directly
implies the corresponding Wick formula for the causal operator-valued
distributions at the level $n$. This is easily shown by the usual 
Wick theorem for ordinary products of Wick monomials,
\begin{equation}
D_n^{\vec{j}}(X,x_n) = \sum_{\vec{s}}  
\< 0 \mid D_n^{\vec{j}+\vec{s}} (X,x_n) \mid 0 \> 
 \frac{ :\Phi^{\vec{s}} (X \cup \{ x_n \}):}{\vec{s}}. 
\end{equation}

This formula reduces the splitting problem of operator-valued
distributions to the splitting of the numerical $C$-number
distributions
\begin{equation}
d_n^r(x_1 - x_n, \ldots, x_{n-1} - x_n) = \<0 \mid D_n^r (X,x_n) \mid 0\>
\end{equation}
The latter only depends on the relative coordinates because of
translational invariance of $D_n$.  Note that the causal support of all
numerical distributions is assured by the fact that they are vacuum
expectation values of operators with causal support. We can construct
well-defined $T_n$ distributions as operators by first splitting the
numerical distributions $d_n^r$ and then by
defining the $T_n$'s as operators using the Wick formula (\ref{Wick}).

The singular behaviour of the distribution $d_n^r$
for $x \rightarrow 0$ is crucial for the splitting problem because 
$\Gamma^+_{n-1} (0) \cap \Gamma^-_{n-1} (0) = \{ 0 \}$.  One therefore has
to classify the singularities of distributions in this region. This
can be characterized in terms 
of the singular order $\omega$ of the 
distribution under consideration    which turns out to be
identical with the usual
power-counting degree \cite{steinmann,fredenhagen1}.
For further details of the theory of
distribution splitting we refer to the literature
\cite{EG,Scharf95} and only make the following {\bf remarks}:\\

$\bullet$ We exclude so-called 'oversubtractions', which correspond
to an increase of the singular
behaviour of the distribution in the splitting process. Thus, we further
specify the splitting problem by requiring in addition 
\be
\label{singular}
\qquad \omega(r) \le \omega(d)\quad \wedge \quad \omega(a) \le
\omega(d).\qquad \mbox{{\bf (D)}} 
\ee
$\bullet$ Moreover, we have to ask whether the splitting solution of a
given numerical distribution $d$ with singular order $\omega(d)$ is
unique. Let $r_1 \in \cal S'$ and $r_2 \in \cal S'$ be two splitting
solutions of the given distribution $d \in \cal S'$. By
construction $r_1$ and $r_2$ have their support in $\Gamma^+$ and
agree with $d$ on $\Gamma^+\setminus \{0\}$, from which follows that $
(r_1 - r_2)$ is a tempered distribution with point support and with
singular order $\omega \le \omega(d):$ \be \mbox{supp}(r_1 - r_2)
\subset \{0\},\quad \omega(r_1-r_2) = \omega(d), \quad (r_1-r_2) \in
\cal S' \ee According to a well-known theorem in the theory of
distributions, we have 
\be \label{freeconstants}
r_1 - r_2 = \sum_{\mid a \mid=0}^{\omega_0} C_a \pa^a\delta(x).  
\ee 
In the case $\omega(d)<0$ which means that $d_n^r$ is regular at the
zero point, the splitting solution is thus unique.  In the case
$\omega(d)\ge 0$ the splitting solution is only determined up to a
local distribution with a fixed maximal singular degree $\omega_0 = 
\omega(d)$. The demands of causality (\ref{GD}) and translational
invariance (\ref{Poi}) leave the constants $ C_a$ in
(\ref{freeconstants}) undetermined. They have to be fixed by
additional normalization conditions.\\
$\bullet$
We want to stress that a normalization ambiguity can already occur in
tree graphs. For example the causal Pauli-Jordan distribution $d_1
:= D(x-y)$ has singular order $\omega(d_1)=-2$, hence $d_2 :=
\partial_\mu^x \partial_\nu^x D(x-y)$ has $\omega(d_2)=0$
(since each derivative increases the singular order by one). 
This implies that the splitting of $d_2$ is not unique according to 
(\ref{freeconstants}). Because the normalization 
ambiguity in tree graphs
will become important in our discussion in the next section let us
discuss this point in more detail.
Note that $ [\phi(x), \phi(y)] = i \hbar D(x-y) = i \hbar ( D^++D^-)$,
where $D^+$ and $D^-$ are the positive, respectively negative
frequency parts of the causal Pauli-Jordan distribution $D$.
In $R'$ and $A'$ of equation (\ref{adva}) the $D^-$ and $-D^+$ occur
in the case of $d_1$.
The so-called natural splitting of the Pauli-Jordan distribution
is given by (see also Appendix A)
\be
\label{natural}
D = D_{ret}-D_{adv}. 
\ee
Here 
$D_{red}$ has retarded support 
and $D_{adv}$ has advanced support. So $r = D_{ret}$  and $a = D_{adv}$
and then $t$ is defined as $t = r - r'  = a - a'$ according to equation 
(\ref{TTT}). This  means in the case under consideration  
$t_1 = D_{red} - D^- = D_{adv} + D^+ = D_F$, so finally we end up 
with the
Feynman propagator. Analogously, the graph with the numerical 
distribution 
$d_2$ leads to a
$t_2$-distribution $t_2 = \partial_\mu^x \partial_\nu^x D_F(x-y) + C
g_{\mu \nu} \delta(x-y)$ with the Feynman propagator $D_F$ and a free
normalization constant $C$ which has to be fixed by a further
condition. \\
$\bullet$ It is important to note that only in tree graphs 
the Feynman propagator $D_F$ occurs.
In loop graphs one gets in $r'$ and $a'$ products of $D^+$ (or $D^-$) 
distributions
which are  well-defined as the direct product of distributions whose
Fourier transform have retarded support ($r'$) and advanced support
($a'$). This does {\it not} lead to
products of Feynman propagators in the $t$ distribution, as it would
be the case if one would use  
the usual Feynman rules that follow from the formal solution 
(\ref{heuristic}). 
For details of explicit splitting solutions in loop graphs we refer 
to \cite{EG,Scharf95}.\\
$\bullet$ We can now discuss the ambiguities on the operator level
using the {\it defining} Wick formula (\ref{Wick}). 
The field itself is included in the
Wick submonomials one starts with, so we have
$[T_1^k(x),\Phi(y)]=0$ if $x \sim y$ 
(i.e. if $x$ and $y$ are spacelike separated) according to condition 
(\ref{caus1}). 
This implies that $T_1^j$ must be in the Borchers
class of the free field $\Phi(x)$ (\cite{epstein2}). It is well-known that the set of Wick
monomials exhausts the Borchers class of a free field. 
This leads to the most general solution of (\ref{GC}).\\
$\bullet$ The Wick formula for time-ordered products (\ref{Wick}) was
used to define the $T_n^j$ distributions 
(including the Wick submonomials) as 
operator-valued distributions. This
formula makes transparent that the normalization ambiguities in $T_n^j$
for different $j$ are not independent. Note that  
the normalization ambiguities of the $T_n^j$'s 
are introduced in this formula through the numerical distributions 
in (\ref{Wick}) only.
Thus, normalization conditions on different $T_n^j$
might lead to a compatibility problem. We already mention here that the
symmetry conditions we analyze in the following only include the
physical $T_n^{j=0}$ distributions, so, no such
compatibility analysis has to be made. \\
$\bullet$
The question of renormalizability naturally arises.
In the EG approach power-counting renormalizable quantum field theories
are the ones where the number of the  constants $C_a$ to be fixed by
physical conditions stays the same to all orders in perturbation
theory. This means that finitely many normalization conditions are
sufficient to determine the $S$-matrix completely.
The latter property, however, does not necessarily mean that this is also
possible when all the symmetry properties of the classical Lagrangian
are maintained, a far more reaching quality generally referred to as
renormalizability. 
As a consequence, power-counting renormalizability is a quality
solely determined by the scaling properties of the theory. In a second
step one tries to prove that there is also a symmetric normalization
of the theory. In particular this holds if the loop normalization 
ambiguity can be fixed in the same 
way as the tree-level normalization ambiguity,
i.e. if the theory is stable under quantum corrections.

If the number of the normalization constants increases with the order 
$n$ of perturbation theory, then the theory is usually called
non-renormalizable. However, if the singular order is bounded 
in every order in perturbation theory then, although the total number of 
physical conditions needed to fix the $S$-matrix completely is infinite, 
this number is finite at each order in perturbation theory and therefore
the theory still has predictive power. Effective field theories belong 
to this class of theories.
One may call these theories `generalized power-counting renormalizable'. 
If in addition the theory is stable under quantum correction then we are 
dealing with `generalized renormalizable' theory.

 Several other properties of the EG formalism are 
discussed in appendix B.

\section{Basic Construction}
\setcounter{equation}{0}

\subsection{The Quantum Noether Condition} 

We shall now present the basic construction of theories with
global(=rigid) and/or local symmetries in the EG formalism. 

As explained in detail in the last section, one starts with a set of free 
fields in the asymptotic Fock space. These fields satisfy their (free) field
equations and certain commutation relations.  To define the theory
one still needs to specify $T_1$, the first term in the $S$-matrix.  
(Actually, as we shall see, even $T_1$ is not free in our 
construction method but is also constrained  by the Quantum Noether Condition).
Given $T_1$ one can, in a well
defined manner, construct iteratively the perturbative $S$ matrix. In
this construction, a finite number of constants (in the case of
a power-counting renormalizable theory (see last section)) remains  
unspecified by the requirements of causality and Poincar\'{e} invariance. 

We are interested in constructing theories where the $S$ matrix is
invariant under a certain symmetry operation generated by a 
well-defined operator $Q$ in the asymptotic Fock space, 
\be 
[Q, S]=0. \label{Sinv} 
\ee 
The operator $Q$ acting on asymptotic fields
generates their asymptotic transformation rules\footnote{
The $(-i\hbar)$ in the right hand side of (\ref{ch1}) as compared 
to (\ref{Qpois}) 
is because in (\ref{Qpois}) we have Poisson brackets whereas in
(\ref{ch1}) quantum commutators.}  
\be 
[Q, \f^A \} = -i \hbar s_0 \f^A, \label{ch1} 
\ee 
where $[A, B\}$ denotes a graded
commutator. The latter are necessarily linear in the asymptotic fields. 
We want to carry out the construction before the adiabatic limit.
Thus, instead of working with (\ref{Sinv}), we shall require 
\be
[Q, T_n(x_1, \ldots, x_n; \hbar) \} = 
\sum_{l=1}^n {\pa \over \pa x^\m_l} T^\m_{n/l}(x_1, \ldots, x_n; \hbar) 
\label{qn} 
\ee
for $n \geq 1$ and for some $T^\m_{n/l}$. We shall often suppress the 
spacetime arguments in the $n$-point functions. We shall also 
use the  abbreviation $\pa/ \pa x^\m_l = \pa^l_\m$.
The meaning of the $T^\m_{n/l}$ will be discussed in detail below.
Equation (\ref{qn}) for $n{=}1$ 
\be 
[Q, T_1 \} = \pa_\m T^\m_{1/1},\label{q1} 
\ee 
imposes restrictions on the starting point of the EG
procedure, namely on the coupling $T_1$.
Once the coupling $T_1$ has
been determined the rest of the equations (\ref{qn}) impose relations
among the constants left unspecified by the requirement of causality
and Poincar\'{e} invariance.  This is analogous to the situation in
the conventional Lagrangian approach where symmetry considerations
restrict the possible terms in the Lagrangian and then the corresponding
symmetries at the quantum level impose certain
relations among the $Z$ factors.

Our considerations apply to the construction of theories with any
global or local symmetry. In the case of linear symmetries, such as
global internal symmetries or discrete $C$, $P$, $T$ symmetries, 
things are much simpler and one does not
need the full machinery developed in this article. This 
is so because linear symmetries can be directly implemented in the 
asymptotic Fock space by means of (anti-)unitary transformations.
To achieve the invariance of the $S$-matrix one only needs to start from a 
coupling $T_1$ invariant under the corresponding linear symmetry.
There is, of course, still the issue of compatibility of the various
symmetries imposed. This question will not be analyzed in this article.

The cases of interest here are non-linear symmetries. In this case, the
asymptotic transformations differ from those of the interacting fields.
Such cases are, for example, the BRST symmetry of gauge theories
and rigid spacetime symmetries such as supersymmetry. In the latter
case, the transformation rules would be linear in the presence of
auxiliary fields.  However, apart from the fact that supersymmetry
auxiliary fields
are not always known, in the EG formalism the fields are on-shell and,
therefore, these auxiliary fields are necessarily absent. 

Since different non-linear transformations may have the same linear
limit it is not {\it a priori} obvious whether a theory constructed by
EG satisfying (\ref{Sinv}) has any underlying non-linear structure
at all. To address this issue one can work out the precise
consequences of the operator equation (\ref{qn}) and 
try to reproduce the Ward identities
derived in the Lagrangian approach using the full non-linear
transformation. This approach has been followed in \cite{DHS95,H95} for
the case of $SU(n)$ gauge theory in the Feynman gauge coupled to
fermions where it was shown that (\ref{qn}) implies the Slavnov-Taylor
identities for connected Green functions. An alternative and
complementary approach is to try to find a direct
correspondence between the Lagrangian approach and the EG formalism.

In the conventional Lagrangian approach the theory is defined by
giving the Lagrangian and specifying a meaningful way to compute
(i.e. regularization/renormalization).  
Our strategy is to identify the Lagrangian within the EG approach. 
If both approaches describe the same theory, then the perturbative 
$S$ matrix should be identical in both. 
The Lagrangian always appears in the $S$-matrix at the tree-level. 
We shall, therefore, identify the Lagrangian with the sum
of $T_1$ and the local terms that arise through tree-level
normalization conditions (notice that in the EG approach one performs 
a perturbative expansion around the free action and not around 
a classical solution of the full theory, so one expects to recover 
the classical Lagrangian through tree-level graphs). 
If this correspondence is correct then, for instance, one 
should be able to understand from the EG point of view why adding a
BRST exact term in the Lagrangian does not change the physics of the
theory.  We will
indeed see that this can be entirely understood using the EG
formalism. If, in addition, one deals with a 
renormalizable theory then loops do not produce any further local 
terms besides  the ones already present in the Lagrangian.
Therefore, the question 
of renormalizability in the Lagrangian 
approach translates, in the 
EG formalism, to the question of whether the local 
normalization ambiguity to all orders reproduces the tree-graph 
normalizations. (The precise definition of these normalization terms
well be given in the next section.)

Let us further remark that these considerations also 
explain why the Lagrangian is such a central 
object in quantum field theory: according to Epstein-Glaser 
the perturbative $S$-matrix is uniquely fixed once one fixes the 
local ambiguity. In a renormalizable theory, the Lagrangian 
precisely fixes this local ambiguity. 

Our proposal for the construction of theories with global and/or local
symmetries in the EG formalism is rather simple. One introduces in
addition to $T_1$ the coupling $g_\m j^\m_0$ in the theory\footnote{
\label{restriction}
The present  considerations do not immediately apply to 
theories that possess asymptotic symmetries
but no associated asymptotic Noether currents. In these cases one still
has a charge $Q$ that generates the symmetry, so
one may still construct these theories using  condition
(\ref{qn}) (see subsection 4.5).  \label{nonoe}}, 
where $j^\m_0$ is the Noether current that generates the 
asymptotic (linear) symmetry transformations.  Actually, as we shall see, 
the coupling $T_1$ itself is determined by the construction. In addition, 
one imposes the condition that ``the Noether current is conserved at the 
quantum level'' (see also (\ref{cond3})), 
\be \label{cons}
\pa_\m \cj_n^{\m} (x_1, \cdots, x_n; \hbar) =0
\ee
where we introduce the notation
\be \label{nota}
\pa_\m \cj_n^{\m} (x_1, \cdots, x_n; \hbar)= 
\sum_{l=1}^n \pa_\m^l \cj_{n/l}^{\m}, 
\ee
and 
\be
\cj^\m_{n/l}=T[T_1 (x_1) \cdots j_0^\m(x_l) \cdots T_1(x_n)].
\ee 
(for $n=1$, $\cj^\m_1(x_1)=j_0^\m(x_1)$).
In other words we consider an $n$-point function
with one insertion of the current $j_0^\m$ at the point $x_l$.
Notice that since the left hand side of (\ref{cons}) is a formal
Laurent series in $\hbar$,
this condition is actually a set of conditions. 
This construction is so natural that one hardly has to motivate it.

We shall show in the remaining of this section that one can construct
using the symmetry condition (\ref{cons}) and the free Noether current
$\cj^\m_1(x_1)=j_0^\m(x_1)$ as a starting point,
any theory with global/local symmetry that can be viewed as deformation
of a free theory (up to restrictions discussed in footnote \ref{nonoe}). 
This class includes all perturbative QFT's.
In addition, we shall establish the equivalence of any theory 
consistently constructed in the EG formalism with a Lagrangian theory
(again up to restriction discussed in footnote \ref{nonoe}).
In the case of interest, namely theories with non-linear symmetry 
transformations, the corresponding Lagrangian can be assumed to be 
obtained by the classical Noether method. I.e. the Lagrangian $\cl$
and the transformation rules $s \f^A$ under which it is invariant,
are both power series in the coupling constant, and are obtained by
solving the equations arising in Noether's method. 
We shall explicitly show that 

\begin{enumerate}
\item the sum of $T_1$ and the tree-level normalizations (see next subsection) 
that arise from 
the requirement (\ref{cons}) coincides with the Lagrangian that
is invariant under the non-linear transformations.  This shows that
the the full non-linear structure is present in the theory,
\item the free Noether's current $j_0^\m$ is renormalized by 
the condition (\ref{cons}) is such a way that it finally generates 
the full non-linear transformations,
\item the loop normalization ambiguity is fixed in the same way 
as at the tree-level one provided the anomaly consistency condition 
has only trivial solutions. This means that 
the theory is then stable under quantum correction,
\item condition (\ref{cons}) is equivalent to  condition
(\ref{qn}).  The latter guarantees the invariance of the $S$-matrix
under the corresponding asymptotic symmetry.
\end{enumerate}
The way $T_1$ and $j_0^\m$ get promoted to the full Lagrangian (point
1) and the full Noether current (point 2), respectively, 
completely parallels the classical Noether method.  However, the EG methods
generates the full quantum theory on the way (point 3), not just a classical
Lagrangian invariant under certain classical symmetry. This motivates
the title of this article.  In particular,  condition (\ref{cons})
(or the equivalent one (\ref{qn})) also contains the symmetry 
constraints at the loop level.  
Points 1 and 2 deal with  condition (\ref{cons}) at tree level
and point 3 covers the loop analysis. Breaking of
(\ref{cons}) by loop corrections corresponds to anomalies. 

There is yet another condition  equivalent to  
(\ref{cons}). 
Let $j^\m_{0,int}$, and $\cl^\m_{1,int}$ 
be the interacting currents corresponding to  $j_0$ and $\cl_1^\m$, 
respectively, constructed 
according to (\ref{defint}) ($\cl^\m_1$ is defined in (\ref{eq1})). 
As it shall be presented in detail \cite{proc}, the Ward identity
\be \label{cond3}
\pa_\m j^\m_{0,int} + \pa_\mu g \cl^\mu_{1, int} =0
\ee
yields the same conditions as (\ref{cons}) 
on the normalization ambiguity of the 
physical correlation functions. In the adiabatic
limit condition (\ref{cond3}) becomes the conservation 
of the interacting current. This is the usual form of the Ward identity
that follows from a symmetry.

\subsection{Off-shell Formulation of the Inductive Hypothesis}

Our goal is to find which are the restrictions on the $T$-products 
implied  by equation (\ref{cons}). Since causality and Poincar\'{e} invariance 
uniquely fix the $T$-products up to local terms, as explained in  
section 3, we only need to discuss the conditions imposed on the local 
normalization ambiguity in (\ref{cons}).
As already mentioned in the introduction, we assume in this paper 
that a consistent and anomaly-free deformation of the asymptotic 
symmetry exists, i.e. we assume that the Quantum Noether method 
works successfully in the cases we consider. This assumption excludes
any true obstructions to the symmetry condition (\ref{cons}) at the tree 
and loop level. A cohomological analysis of possible true 
obstructions of 
condition ({\ref{cons}) without using the quantum action principle
will be presented in a separate paper \cite{paper2}.

We shall follow an iterative approach following the inductive 
EG construction. 
Namely, we shall 
assume that (\ref{cons}) is satisfied 
for all $m<n$, and then we shall examine the conditions implied
by (\ref{cons}) at $n$th order. According to section 3 this involves
three steps:  We first 
construct the corresponding causal distribution $D_n [j_0\ T_1 \cdots
T_1]$, then we have to split $D_n$  to obtain 
$T_{c,n} [j_0\  T_1 \cdots T_1]$, and finally we impose (\ref{cons})
that leads to conditions on the normalization ambiguity of $T_n$.  
The notation $T_c$ indicates that we use the natural splitting
solution (i.e. the Feynman propagator
is used in tree-graphs, see (\ref{natural})) in tree 
graph contributions. The latter is our reference solution. When we refer 
to local normalization terms  in tree graph contributions in EG they
are  always defined with respect to $T_c$. 
Points 1 and 2 crucially depend on this choice. There is a good reason, 
however, why this is what one should do: Only with natural splitting 
in tree-level graphs the contraction between two fields 
becomes equal to the Feynman propagator. As we already argued, 
we shall identify the Lagrangian with local terms in tree-level graphs in 
the $S$-matrix. In the Lagrangian approach these graphs have been 
constructed using Feynman propagators. So, in order to compare the two 
approaches one has to use the natural  splitting solution.
It is only for the sake of  comparison that the natural  splitting
solution becomes distinguished. In the analysis of (\ref{cons}) at the 
loop level one may likewise choose a reference splitting solution. 
In this case, however, there is no `preferred' reference solution, but also 
no need to explicitly specify one. In the following the subscript $c$
will denote natural splitting in tree-graphs and some fixed  
reference splitting in loops. 

Let us start by noting that having satisfied our fundamental 
Quantum Noether condition (\ref{cons}) for all $m<n$, namely
\be \label{cons2}
\pa_\m \cj_m^{\m} (x_1, \cdots, x_m; \hbar)= 
\sum_{l=1}^m \pa_\m^l \cj_{m/l}^{\m} = 0,
\quad \forall m<n,  
\ee
then  equation (\ref{cons}) at the $n$th order 
can be violated by a local distribution $A_n(\hbar)$
(which we shall call anomaly term) only:
\be \label{cons3}
\pa_\m \cj_n^{\m} (x_1, \cdots, x_n;\hbar)= \sum_{l=1}^n 
\pa_\m^l \cj_{n/l}^{\m} = A_n(\hbar),
\ee
Let us give a short proof of the latter statement: From 
(\ref{cons2}) one can derive the 
analogous condition for the causal distribution $\cd$ at 
the $n$th order: 
\be
\label{cons4}
\sum_{l=1}^n \pa_\m^l \cd^\n_{n/l} =0,
\ee
where $\cd^\m_{n/l}$ denotes $\cj^\m_{n/l}$ at the $D$-level. 
The latter step 
is somehow trivial using the basic formulae (\ref{ret})-(\ref{adva}) 
which involve only tensor products of the known $T_m$ 
products with $m<n$ which fulfil (\ref{cons2}). Knowing (\ref{cons4}), 
we have to split the causal distributions $\cd^\n_{n/l}$:  
Since the splitting solution $\car_{n/l}$ of $\cd_{n/l}$ 
fulfils $\car_{n/l}=\cd_{n/l}$ on $\Gamma ^+ \setminus 
\{(x_n,...,x_n)\}$ and $\car_{n/l}=0$ on $(\Gamma ^+)^c$, 
the symmetry condition  can be violated in this process 
only in the single point $(x_n,...,x_n)$, i.e. 
by local terms. 
Therefore,  condition (\ref{cons}) at $n$th order can only be
violated by local terms, denoted by $A_n(\hbar)$ in (\ref{cons3}). 
The anomaly terms $A(\hbar)$ are a formal Laurent series in 
$\hbar$ since the left hand 
side in (\ref{cons3}) is. In addition, they are restricted by the power 
counting condition (\ref{singular}). Notice that we allow for theories
with different (but finite) maximal singular order $\omega$ 
at every order in perturbation theory (i.e. we consider the class of 
`generalized renormalizable' theories, see section 3). 

We shall now present an off-shell version\footnote{
By off-shell we mean that we relax the field equations
of the fields $\phi^A$.} of the inductive hypothesis.
The assumption that the Quantum Noether method works successfully
means that there exist local normalizations such that 
(\ref{cons2}) is satisfied when the field equations are satisfied. 
This does not mean, however, that (\ref{cons2}) 
is satisfied when any splitting is used.
Actually, generically after natural splitting 
(this refers to tree-level graphs, for loop graphs one uses some
reference splitting solution) one ends up with 
\be
\pa_\m \cj_{c,m}^{\m} = A_{c,m},
\ee
where the subscript $c$ indicates that the natural splitting has been 
used at tree-graphs. 
Our assumption only means that the anomaly $A_{c,m}$ is a divergence 
up to terms $B_m$ that vanish when the free field equations are used, i.e.
\be \label{an1}
A_{c,m} = \pa_\m A_{c,m}^\m + B_m
\ee
where $A_{m,c}^\m$ and $B_m$ are some local distributions (since $A_{c,m}$ 
is local). This decomposition is not unique since one can move derivatives 
of field equation terms from $B_m$ to $A_{c, m}^\m${}
We fix this freedom
by demanding that $B_m$ does not contain any derivatives of field equations.
Let us show this explicitly.
To derive the general form of $B_m$ we first note that 
it should have the general form of a local distribution\footnote{
We use the following abbreviations for the delta function distributions
$\d^{(m)}=\d(x_1, \ldots, x_m)=\d(x_1-x_2)\cdots\d(x_{m-1}-x_m)$. 
}
\be \label{ansatz1}
B_m = O_{1,m} \delta^{(m)} + \pa_\mu O^\m_{2,m}, \qquad 
O^\m_{2,m}=\sum O_{2,m}^{\m \a_1 \cdots \a_n} \pa_{\a_1} \d(x_1-x_n)\cdots
\pa_{\a_n} \d(x_{n-1}-x_n),
\end{equation}
which is easy to prove by recursion on the number of derivatives 
and integration by parts.
Notice that because the power counting degree is bounded at each order
in perturbation theory the series terminates 
after a finite number of terms. The operators $O_{i,m}, i=1,2,\ldots$, are 
in general unrestricted, but in our case they should be such that $B_m$ 
vanishes when the free field equations are satisfied. This means in 
particular that $O_{1,m}$ has the form
\be \label{o1}
O_{1,m} = S^{A;m} \ck_{AB} \f^B 
+ \sum_p S^{A;m}_{\m_1 \cdots \m_p} \pa^{\m_1} 
\cdots \pa^{\m_p} \ck_{AB} \f^B
\ee
where 
\be
\ck_{AB} \f^B= \pa^\m {\pa \cl_0 \over \pa (\pa^\m \f^A)} 
- {\pa \cl_0 \over \pa \f^A}   
\ee
are the free field equations. The fact that $O_{1,m}$ is linear
in the field equations follows from the fact that the field equations
are created because we act with a derivative on a T-product with 
one current insertion. One derivative can only create one field
equation. 

Remembering that (\ref{ansatz1}) is a distributional relation, one may 
integrate by parts the derivatives from the field equations to obtain
\be \label{newo1}
O'_{1,m}=(S^{A;m} 
+ \sum_p (-1)^p \pa^{\m_1} \cdots \pa^{\m_p} S^{A;m}_{\m_1 \cdots \m_p})
\ck_{AB} \f^B,
\ee
and appropriate modifications of the 
$O_{i,m}, i>0$.  Notice now that one may always factor out a derivative from 
the terms involving the $O_{i,m}, i>0$. This means that these terms 
can be moved into $A^\m_{c, n}$. The latter is finally 
removed by appropriately
fixing the local normalization freedom of the left hand side. 
Let us also define
\be
R^{A;m} = S^{A;m} 
+ \sum_p (-1)^p \pa^{\m_1} \cdots \pa^{\m_p} S^{A;m}_{\m_1 \cdots \m_p}
\ee

An additional ambiguity is related to the global symmetries of the 
free action. If one makes the transformation
\be \label{amb}
A_{c, m}^\m \to A_{c, m}^\m + \tilde{j^\m}; \quad
R^{A;m} \to R^{A;m} - \tilde{s} \f^A 
\ee
where $\tilde{j^\m}$ is a Noether current that generates the symmetry 
transformations $\tilde{s} \f^A$, then the right hand side of (\ref{an1})
remains unchanged. To fix this ambiguity we demand that $R^{A;m}$
do not contain any summand which is itself a symmetry transformation 
of the free action. (In practice, one would never have to deal with 
this problem unless one does by hand the substitutions (\ref{amb})). 

In this manner we are lead to the following off-shell 
representation of the inductive hypothesis: for $m<n$,
\be \label{tfeq}
\sum_{l=1}^m \pa^l_\mu \cj_{m/l}^{\m} =
\sum_{A} R^{A;m}(\hbar) \ck_{AB} \f^B \d(x_1, \ldots, x_m). 
\ee 
The coefficients $R^{A;m}(\hbar)$ may, in general, receive tree and loop 
contributions. We shall show below that this off-shell representation
provides an alternative and simplified way of obtaining  
local terms arising from tree-level graphs. 

We first concentrate on analyzing the condition (\ref{cons}) 
at tree-level. We shall consider the loop case afterwards.
We therefore only need the $\hbar^0$ part of (\ref{tfeq}).
Let us define 
\be
\label{delta}
s_{(m-1)}\f^A = {1 \over  m!} R^{A;m}(\hbar^0); \ \ m > 1,
\ee 
(we shall see below that this formula also holds for $m=1$).
Depending on the theory under consideration the quantities $R^{A;m}(\hbar^0)$
may be zero after some value of $m$. Without loss of generality we 
assume that they are zero for $m>k+1$, for some integer $k$ (which 
may be infinity; the same applies for $k'$ below.).
We shall show below that 
\be
s \f^A = \sum_{m=0}^k g^m s_m \f^A
\ee
are symmetry transformation rules that leave the Lagrangian invariant
(up to total derivatives)
\be \label{lagr}
\cl = \sum_{m=0}^{k'} g^m \cl_m,
\ee 
where $k'$ is also an integer (generically not equal to $k$).
The Lagrangian $\cl$ will be determined from the tree-level normalization
conditions as follows,
\be \label{lagdef} 
\cl_m = {\hbar \over i} {N_m \over m!}, \quad {\rm for} \quad m>1,
\ee
where $N_m$ denotes the local normalization ambiguity of  
$T_m[T_1(x_1)...T_1(x_m)]$ in tree graphs defined with respect 
to the naturally split solution. For $m=1$, $\cl_1=(\hbar/i)T_1$.
The factor $m!$ reflects the fact that 
$T_m[...]$ appears in (\ref{GC}) with a combinatorial 
factors $m!$ while the factor $\hbar/i$ is there to cancel the 
overall factor $i/\hbar$ that multiplies the action in the 
tree-level $S$-matrix. Notice that we regard 
(\ref{lagdef}) as definition of $\cl_m$. 

\begin{sloppypar}
To understand how the off-shell formulation simplifies the 
calculation of local terms arising from tree-level graphs
we start by first describing the traditional way to do  such a 
calculation. 
In order to obtain the local terms, 
one first constructs 
 $T_{c,n}[j^\mu_0(x_1)T_1(x_2)...T_1(x_n)]$, differentiates with respect 
to the variable   of the current and symmetrizes in all variables.
$T_{c,n}[j^\mu_0(x_1)T_1(x_2)...T_1(x_n)]$ 
involves many terms and there will be a large number of cancellations
after differentiating and symmetrizing.
In particular, we already know from 
equation (\ref{cons3}) that all non-local
terms will cancel among themselves.
So the idea (which gets implemented with the help 
of the off-shell formulation) is to only concentrate 
on possible local terms anticipating the 
cancellation of all non-local terms.
\end{sloppypar}

{}From  (\ref{ret}), (\ref{adva}) we know 
that the causal distribution $D_n[j_0^\m(x_1) T_1(x_2) ... T_1(x_n)]$ 
is equal to the sum of terms 
which are products of $T$-products constructed at lower orders 
one of which
contains $j_0^\m$ as a vertex.
To calculate $T_{c,n}[j_0^\m(x_1) T_1(x_2)...T_1(x_n)]$ one 
first does all contractions in $D_n$ and then splits the solution.
Each tree-level contraction between two fields $\f^A$ and $\f^B$ 
yields after 
natural splitting a factor  $\cd^{AB}$, 
where $\cd^{AB}$ is the inverse of the corresponding kinetic operator 
$\ck_{AB}$, 
\be \label{inve}
{i \over \hbar} \ck_{AB} \cd^{BC}=\d_A^C 
\ee
where $\d_A^C$ contains a delta function, and we have included also the 
$i$ and $\hbar$ factors (see appendix A). 
Let us also denote by $\co^\m_{AB}$ the operator
associated with the fields $\f^A, \f^B$ that satisfies 
$\pa_\m \co_{AB}^\m = \ck_{AB}$.
The expression for $T_{c,n}[j_0^\m(x_1) T_1(x_2)...T_1(x_n)]$ 
will contain, among other terms, terms of the form
\footnote{In (\ref{mec1}),
as well as in later formulae, Wick-ordering is always understood.
We also suppress two delta distributions that set $n$ of the variables
equal to $x$ and the remaining $n-m$ equal to $y$. In order to keep 
the notation as simple as possible we shall often suppress such delta 
distributions. In all cases one may insert these delta distribution
by simple inspection of the formulae.} 
\be \label{mec1}
 S^1 (x)  S^2 (y)  \co^\m_{AB}(x)  \cd^{BC}(x-y),
\ee 
where $x$ is the variable of the current and $S^1, S^2$ 
are local terms  at order $m \ (m<n)$,
respectively, $n{-}m$.  
Upon differentiating with $\pa_\m^{x}$  this term will 
bring among other terms the local term
\be \label{local}
S^1 S^2  \d_A^C \d(x-y). 
\ee
(we have now explicitly written the delta function to emphasize that 
this is a local term).
The mechanism we just described is the only one
that creates local terms out of tree-level graphs.
Diagrammatically, after we push the derivative in we get an inverse 
propagator that cancels the propagator between $x$ and $y$, thus,
rendering the graph local. The fact that $S^1$ and $S^2$ are
local is, of course, essential. Local terms proportional 
to derivatives of the $\delta$ distribution
are constructed in an analogous  way.

\begin{sloppypar}
The term (\ref{mec1}) originated from the following $T$-product
\be \label{mec2}
T_{c,n} [ (S^1 \co^\m_{AB} \f^B\d^{(m)})(x) (S^2 \f^C)\d^{(n-m)}(y)] , 
\ee
upon contraction between $\f^B$ and $\f^C$. We are ultimately interested
in computing $\pa_\m T_{c,n}[j_0^\m(x_1) T_1(x_2)...T_1(x_n)]$,
so we need to compute 
$\pa_\m T_{c,n}[(S^1 \co^\m_{AB} \f^B \d^{(m)})(x) (S^2 
\f^C \d^{(n-m)})(y)]$.
Moving the derivative inside the correlation function we get 
(among other terms)
\be \label{mec3}
T_{c,n}[(S^1 \ck_{AB} \f^B)(x) (S^2 \f^C)(y)]
\ee
Upon contracting the $\f^B$ with a $\f$ in $(S^2 \f^C)$ 
one obtains a local term. In particular, the local term in (\ref{local}) 
is simply obtained from the contraction between the explicit $\f$'s
(i.e. the $\f^B$ and $\f^C$) in (\ref{mec3}). We shall call this 
kind of contractions, namely the ones that involve the field
$\f^B$ in the field equation $\ck_{AB} \f^B$, ``relevant contractions''.
These yield the local terms that one obtains by doing the calculation 
in the ``traditional way''. All other contractions (``irrelevant
contractions'') 
generically yield non-local terms proportional to field equations.
These are not relevant in our case and they will be discarded.
\end{sloppypar}

In this way we are led to an 
alternative and more systematic way of obtaining all tree-level local
terms. One first differentiates and then does the ``relevant contractions''. 
However, in doing the calculation in this
way one should {\it not} use the field equations 
before the end of the calculation. The terms that are 
proportional to the field
equations are the source of the local terms. 

Let us now make contact with the off-shell formulation
of the induction hypothesis. After differentiation, the causal 
distribution 
$\sum_{l=1}^n \pa_\m^l \cd^\m_{n/l}=0$ at the $n'$th order
consists of a sum of terms each of these being  
a tensor product of 
$T_m[T_1 ... T_1 \partial{\cdot}j_0 T_1 ... T_1)$ ($m<n$) with 
$T$-products that involve 
only $T_1$ vertices according to the general formulae
(\ref{ret},\ref{adva},\ref{D-dist}).
By the off-shell induction hypothesis, we have for all $m<n$ 
\be \label{offshell}
\sum_{l=1}^m \pa^l_\mu \cj_{m/l}^{\m} =
\sum_{A} (m! s_{m-1} \f^A)  \ck_{AB} \f^B \d^{(m)}. 
\ee
At order $n$ the ``relevant contractions'', namely the
contractions between the $\f^B$ in the right hand side of (\ref{offshell})
and $\f$ in local terms, yield the sought-for local terms.
This implies, in particular, that no local term arises from
terms in $D_n$ that are products of more than two $T$ products.
This is in accordance with the diagrammatic picture of creation 
of local terms that we mentioned above.
In this manner we get the following general formula for the 
local term $A_{c,n}$  arising through tree-level contractions at level
$n$,
\be \label{loc}
A_{c,n}(tree) = \sum_{\pi \in \Pi^n} \sum_{m=1}^{n-1}
\pa_\m \cj_m^\m(x_{\p(1)}, \ldots, x_{\p(m)}) 
N_{n-m}\d(x_{\p(k+1)}, \ldots, x_{\p(n)})
\ee
where it is understood that in the right hand side only 
``relevant contractions'' are  
made. The factors $N_{n-m}$ are tree-level normalization terms of 
the $T$-products that contain $n-m$ $T_1$ vertices.

\subsection{Analysis of the Quantum Noether Condition at Tree-level}

In this section we  analyze the formula (\ref{loc}) for all $n$.

For $n=1$, we have $\cj_1^\m=j_0^\m$. Then from (\ref{jdiv}) it follows that 
\be \label{n1}
\pa_\m \cj_1^\m = s_0 \f^A K_{AB} \f^B.
\ee
Therefore, $R^{A;1}=s_0 \f^A$ as we promised.

Let us now move to the $n=2$ case. From our previous 
discussion follows immediately that 
\be
\pa^{x_1}_\m T_{c,2}[j^\m_0(x_1) \f^A(x_2)] = {\hbar \over i} 
s_0 \f^A \d(x_1-x_2)
\ee
where the factor $(\hbar/i)$ originate from the contraction
(see (\ref{inve})).
Using the derivation property of single Wick contractions 
we immediately
get that for any local function $f(\f^A)$ of the fields 
that do {\it not}
contain derivatives of the fields a similar relation holds,
\be
\pa^{x_1}_\m T_{c,2}[j^\m_0(x_1) f(\f^A)(x_2)] = {\hbar \over i} 
s_0 f(\f^A) \d(x_1-x_2).
\ee
Let us know consider derivative terms. In this case,
\be
\label{n2}
\pa^{x_1}_\m T_{c,2}[j^\m_0(x_1) (\pa_k \f^A)(x_2)] = {\hbar \over i}
(s_0 \f^A)(x_1) 
\pa_k^{x_2} \d(x_1-x_2)
\ee  
Symmetrizing\footnote{
The current $j_0^\m$ contains the parameter of the transformation rule
and is therefore bosonic. If one were to use the current without the 
parameter then one would need to graded-symmetrize in $x_1$ and $x_2$.}
this expression with respect to $x_1$ and $x_2$ and 
using the distributional identity 
\be
\label{distr}
a(x_1) \pa_{x_2} \d(x_1-x_2) + a(x_2) \pa_{x_1} \d(x_1-x_2) = 
(\pa a) \d(x_1-x_2)
\ee
we get 
\be
\pa^{x_1}_\m T_c[j^\m_0(x_1) (\pa_k \f^A)(x_2)]
+ \pa^{x_2}_\m T_c[(\pa_k \f^A)(x_1) j^\m_0(x_2)] = {\hbar \over i}
\pa_k (s_0 \f^A) 
\d(x_1-x_2).
\ee
Notice that the corresponding relation with $(\pa_k \f^A)$ 
replaced by $\f^A$ has an extra factor of $2$ in the right hand side. 
Combining these results we obtain\footnote{In this article we 
consider only theories that depend arbitrarily on $\f^A$ and 
its first derivative $\pa_\m \f^A$. 
One could also study theories
that depend on higher derivatives of $\f^A$, $\pa_{m_1}...\pa_{m_p} \f^A$,
for some integer $p$. However, it is not clear whether such theories 
have any relevance to the physical world as classically they have
no phase space and therefore it is unclear how to
canonically quantize them. Notice however that one may construct them 
in the EG approach. We will not consider
these theories in this article.} (we use the notation 
introduced in (\ref{nota}))
\be \label{n=2}
\pa_\m \cj^{\m}_{c,2} (x_1, x_2) = {\hbar \over i}
(2 s_0 T_1 - \pa^\m (\frac{\pa T_1}{\pa (\pa_\m \f^A)} s_0 \f^A)) 
\d(x_1 - x_2) = A_{c,2}
\ee 
Inserting the definition of $\cl_1=(\hbar/i) T_1$ the right hand side 
of (\ref{n=2}) becomes real and independent of $\hbar$. 

Our objective is to fix the tree-level normalization freedom 
with respect to $\cj^{\m}_{c,2}$
such that (\ref{cons}) holds (when the free field equations 
are satisfied). This is possible if and only if
\be \label{eq1}
s_0 \cl_1 = \pa_\m \cl_{1}^\m + \half B_2
\ee
for some  $  \cl_{1}^\m$ and $B_2$, where 
$B_2$ vanishes when the free field equations are satisfied 
(the factor $1/2$ has been inserted such that we agree with (\ref{an1})). 
By our assumption-the Quantum Noether Method works successfully-
there is a pair $(\cl_1,\cl_1^\m)$ which solves (\ref{eq1}).  
We emphasize that also  $\cl_1=(\hbar/i) T_1$ is not an input in
 our  construction method but also is determined by the Quantum
Noether condition (\ref{cons}). So only the free Noether current
$j_0^\m$ is used in the  defining equation of the Quantum Noether method.
Going back to (\ref{n=2}) we obtain 
\bea
\pa_{\m} \cj^{\m}_{c,2}(x_1, x_2)&=& 
2! (s_1 \f^A) \ck_{AB} \f^B \d(x_1 - x_2)\nonu
&&- \pa_\m (-2\cl_{1}^\m
+ \frac{\pa \cl_1}{\pa (\pa_\m \f^A)} s_0 \f^A) \d(x_1 - x_2)
\eea
According to (\ref{delta}) the latter equation defines $s_1$.

Now we consider the normalization ambiguity of  $T_2[j_0 T_1]$,
\be \label{defj1}
T_2[j_0^\m(x_1) T_1(x_2)]=T_{c,2} [j_0^\m(x_1) T_1(x_2)] + 
j_1^\m \d(x_1-x_2)
\ee
Demanding that 
\be
\label{n22}
\pa_\m \cj^{\m}_{2}(x_1, x_2) = 2! s_1 \f^A \ck_{AB} \f^B \d(x_1-x_2)
\ee
we obtain
\be \label{j1}
j_1^\m = \frac{\pa \cl_1}{\pa (\pa_\m \f^A)} s_0 \f^A  
-2\cl_{1}^\m.
\ee

The discussion for $3 \leq n \leq k+1$ goes the same way as the $n=2$
case. However, it is instructive to also directly work out the $n=3$ 
case as it is still relatively easier than the general case 
but sufficiently
more complicated than the $n=2$ case. 
Using (\ref{ret}) and (\ref{adva}) for $n=3$ and 
afterwards naturally splitting one gets 
\be 
\label{n3}
\pa_\m \cj^{\m}_{c, 3}(x_1, x_2, x_3) =
\pa_\m \cj^{\m}_{2}(x_1,x_2) T_1(x_3)
+\pa_\m \cj_1^\m(x_1) N_2(x_2, x_3)
+ {\rm cyclic\ in}\ x_1, x_2, x_3, 
\ee 
(this is equation (\ref{loc}) for $n=3$). 
In writing this expression we have discarded 
all terms on the right hand side 
that do not contribute any local terms.  In particular, it is 
understood that 
only ``relevant contractions'' are made. $N_2$ denotes the 
tree-normalization
term of $T_2$ which is uniquely defined with respect to $T_{c,2}$.

Using our previous results (\ref{n1}),(\ref{n22})
and remembering that the derivative terms
should be treated with care we obtain 
\bea \label{n=3} 
\pa_\m \cj^{\m}_{c, 3}(x_1, x_2, x_3) &=&
[3! (s_1 \cl_1 + s_0 \cl_2) \nonu
&-& 2! \pa_\m ( 2 \frac{\pa \cl_1}{\pa(\pa_\m \f^A)} s_1 \f^A
+ \frac{\pa \cl_2}{\pa(\pa_\m \f^A)}s_0 \f^A)] \d (x_1, x_2, x_3)
\eea
where we have used the definition in (\ref{lagdef}).
The Quantum Noether condition (\ref{cons}) is satisfied if and only if 
\be \label{noether3}
s_1 \cl_1 + s_0 \cl_2 = \pa_\m \cl_2^\m + s_2 \f^A \ck_{AB} \f^B
\ee
for some  $\cl_2^\m$ and $ s_2 \f^A$.
Substituting back in (\ref{n=3}) we obtain 
\bea
\pa_\m \cj^{\m}_{c, 3}(x_1, x_2, x_3) &=&
3! (s_2 \f^A) \ck_{AB} \f^B \d (x_1,x_2,x_3) \nonu
\hspace{-3.7cm}&-&\!\!\!\!\pa_\m [-3!\cl_2^\m 
+ 2!(\frac{\pa \cl_2}{\pa(\pa_\m \f^A)}s_0 \f^A
+2\frac{\pa \cl_1}{\pa(\pa_\m \f^A)}s_1\f^A)]\d (x_1, x_2, x_3)
\eea
In a similar way as before we define 
\be \label{j2def}
T_3 [j_0^\m(x_1) T_1(x_2) T_1 (x_3)]=
T_{c,3} [j_0^\m(x_1) T_1(x_2) T_1(x_3)] 
+ j_2^\m \d(x_1, x_2, x_3)
\ee
Then 
\be
\pa_\m \cj^{\m}_{3}(x_1, x_2, x_3) = 3! s_2 \f^A \ck_{AB} \f^B 
\d(x_1, x_2, x_3)
\ee
provided 
\be \label{j2}
j_2^\m= -3! \cl_2^\m
+ 2!(\frac{\pa \cl_2}{\pa(\pa_\m \f^A)} s_0 \f^A
+2 \frac{\pa \cl_1}{\pa(\pa_\m \f^A)} s_1 \f^A) 
\ee
The general case for $ 2 \leq n \leq k+1$ can be worked 
out in a completely analogous
way. The result for the current is 
\be \label{jn}
j_{n-1}^\m= -n!\cl_{n-1}^\m
+(n-1)! \sum_{l=0}^{n-2} (l+1) \frac{\pa \cl_{n-1-l}}
{\pa(\pa_\m \f^A)} s_l \f^A 
\ee
where we have used the definition in (\ref{lagdef}).
To derive this result one may use following distributional identity
\bea \label{ident}
&&\sum_{\p \in \P^n} \d(x_{\p (1)}, x_{\p (2)}, \ldots, x_{\p (k)})
A \pa^\m_{x_{\p (k)}} (B  \d(x_{\p (k)}, x_{\p (k+1)}, \ldots, 
x_{\p (n)})= \nonu 
&&[\left(\begin{array}{c} n \\  k  \end{array} \right)
A \pa^\m B 
- \left(\begin{array}{c} n-1 \\  n-k-1 \end{array} \right) \pa^\m (AB)] 
\d(x_1, \ldots, x_n)
\eea
where $\left(\begin{array}{c} n \\  k \end{array} \right)$ 
is the binomial coefficient. In addition, 
\be
\pa_\m \cj^{\m}_n = n! s_{n-1} \f^A \ck_{AB} \f^B \d(x_1, \ldots, x_n)
\ee

Let us now move to the $n > k+1$ case. (In the case of ``generalized
renormalizable'' theories only the analysis of the case $n \leq k+1$ 
is present since $k$ tends to infinity). One gets 
\bea
\label{higher}
\pa_\m \cj^{\m}_{c,n}&=&
n![s_0 \cl_{n-1} + s_1 \cl_{n-2}
+ \cdots + s_k \cl_{n-1-k} ] \nonu
&&-(n-1)! \pa^\m \sum_{l=1}^{k} 
l \frac{\pa \cl_{n-l}}{\pa(\pa_\m \f^A)} s_{l-1} \f^A
\eea
where again use of (\ref{ident}) has been made. 
This equation now implies that 
\be \label{n>k}
s_0 \cl_{n-1} + s_1 \cl_{n-2} 
+ \cdots + s_k \cl_{n-1-k}=
\pa_\m \cl_{n-1}^\m.
\ee
One achieves 
\be 
\pa_\m \cj^{\m}_{n}=0
\ee
by renormalizing the current 
\be \label{jn1}
j_{n-1}^\m=-n!\cl^\m_{n-1} + (n-1)! \sum_{l=1}^{k} 
l \frac{\pa \cl_{n-l}}{\pa(\pa_\m \f^A)} s_{l-1} \f^A
\ee
and without the need to use the free field equations. 
Depending on the theory under consideration the $\cl_n$'s will 
be zero for $n>k'$, for some integer $k'$. Given the integers $k$ and
$k'$, there is also an integer $k''$ (determined from the other two)
such that $\cl^\m_n=0$, for $n>k''$. 

Let us recapitulate. We have calculated all local terms 
that arise from the tree level diagrams. Summing up the 
necessary and sufficient conditions (\ref{eq1}), (\ref{noether3}), 
(\ref{n>k}) for the Quantum Noether method to 
hold at tree level we obtain,
\be
s \sum_{l=1}^{k'} g^l \cl_l = \sum_{l=1}^{k''} \pa_\m \cl_l^\m 
+ (\sum_{l=1}^k g^l s_l \f^A) \ck_{AB} \f^B
\ee
Using $s_0 \cl_0 = \pa_\m k^\m_0$ and for $l \leq k$
\be
s_l \f^A \ck_{AB} \f^B = \pa_\m ({\pa \cl_0 \over \pa(\pa_\m \f^A)} s_l \f^A)
-s_l \cl_0
\ee
we obtain,
\be \label{treecon}
s \cl = \pa_\m (\sum_{l=0}^{k''} g^l k_l^\m)
\ee 
where, for $1<l \leq k$, 
\be \label{kdef}
k_l^\m = \cl_l^\m + {\pa \cl_0 \over \pa(\pa_\m \f^A)} s_l \f^A
\ee
and for $l>k$, $k_l^\m = \cl_l^\m$.
We therefore find that $\cl$ is invariant under the symmetry 
transformation,
\be
s \f^A = \sum_{l=0}^k g^l s_l \f^A.
\ee
According to Noether's theorem there is an associated  Noether current
given by (\ref{noether}). Using (\ref{kdef}) one may check that 
the current normalization terms $j_m^\m$ ((\ref{j1}), (\ref{j2}),
(\ref{jn}), (\ref{jn1})) are in one-to-one correspondence
with the terms in the Noether current
(compare with (\ref{noether})); 
they only differ by combinatorial factors related to the 
perturbative expansion. Therefore the current $j_0$ indeed renormalizes to the 
full non-linear current. This finishes the proof of points 1 and 2.

\subsection{Analysis of the Quantum Noether Condition at Loop-level} 

We now move to point 3 and consider what happens in loops.
If the local normalization ambiguity
to all orders in $\hbar$ reproduces the tree-graph normalizations
then we are dealing with a (generalized) renormalizable theory.
We have seen that the tree level analysis leads to  formula
(\ref{loc}) for the local terms at order $n$. At higher loop 
level this formula is not correct any longer. In addition to the 
tree level terms present in (\ref{loc}) one also has loop graphs
between the correlation function that contains the $j_0^\m$ 
vertex and the rest. Furthermore, the transformation rules 
may receive quantum correction, i.e. $R^{A;m}$ is a series in $\hbar$. 
Symbolically one has
\bea \label{loop}
A_{c,n}(\hbar^M)&=&\sum_{m_1+m_2=n}\{ 
(R^{A;m_1}(\hbar^0) \ck_{AB} \f^B \d^{(m_1)} N_{m_2}(\hbar^{M}) \d^{(m_2)})
\nonu
&+& \sum_{M_1+M_2+M_3=M} 
[R^{A;m_1}(\hbar^{M_1}) N_{m_2}(\hbar^{M_2})](M_3 \ \ loops)\}
\eea
where $N_n(\hbar^M)$ denotes the local normalization freedom of 
$T_n[T_1(x_1)...T_1(x_n)]$ in the $M$-th loop level. 
In the first line in the right hand side of (\ref{loop}) 
tree-level ``relevant contractions'' are understood. These terms correspond
to the right hand side of (\ref{loc}). The second line in (\ref{loop})
contains the new terms on top of the ones present in (\ref{loc}). 
We shall collectively denote these terms $L_n$. 

The analysis of the first line in the  right hand side of (\ref{loop})
is exactly the same as the tree-level analysis. The loop terms $L_n$
in (\ref{loop}) are also local terms. Their general form is therefore
of the form (\ref{ansatz1}),
\be \label{ansatz2}
L_n = n! L_{1,n} \delta^{(n)} + \pa_\m L_{2,n}^{\mu}, 
\qquad 
L^\m_{2,m}=\sum L_{2,m}^{\m \a_1 \cdots \a_n} \pa_{\a_1} \d(x_1-x_n)\cdots
\pa_{\a_n} \d(x_{n-1}-x_n),
\end{equation}
(the $n!$ in the first term was added for later convenience).
The term $L_{2, n}$ is a total derivative term.
We remove it by appropriately fixing the local 
normalization freedom of the current correlation function.
Hence, we obtain
\be
A_{n,c}(\hbar^M)=n![(s_0 \cl_{n-1}(\hbar^M) + \cdots 
+ s_k \cl_{n-1-k}(\hbar^M)) 
+ L_{1, n} (\hbar^M)] \d^{(n)} 
\ee
where we have extended the definition (\ref{lagdef}) to cover also 
the loop case. From our assumption that the Quantum Noether method works 
successfully there follows
\be  
s_0 \cl_{n-1}(\hbar^M) + \cdots 
+ s_k \cl_{n-1-k}(\hbar^M) 
+ L_{1, n} (\hbar^M)
= \pa_\m C_n^\m(\hbar^M)
\ee
for some $C_n^\m$. Summing up these relations we obtain,
\be \label{stab}
s \cl(\hbar^M) + L_1 (\hbar^M) = \pa_\m C^\m(\hbar^M)
\ee
where we have defined $L_1 = \sum_n L_{1, n}$, $C=\sum_n C_n$ and we 
have extended the definition (\ref{lagr}) at the loop level.
Equation (\ref{stab}) constrains the local terms 
$N(\hbar^M)(=\sum_n N_n(\hbar^M))$.
Notice that $L_1$ only depends on $N(\hbar^K)$ for $K<N$ (this follows from 
simple $\hbar$ counting, see (\ref{loop})). A sufficient and necessary 
condition for loop terms to satisfy the same condition as the tree-level 
terms is 
\be \label{exact}
L_1(\hbar^M) = s M(\hbar^M) + \pa_\m M^\m(\hbar^M)
\ee
If this relation is satisfied then the renormalized local terms
$\cl'(\hbar^M)=\cl(\hbar^M) + M(\hbar^M)$ satisfy
\be
s \cl'(\hbar^M) = \pa^\m C'_\m
\ee
where $C'_\m = C_\m - M_\m$, i.e. the same equation as (\ref{treecon})
satisfied by the tree-level normalizations, and the theory is stable.

If we are considering a BRST-like symmetry then (\ref{exact}) will 
always be satisfied if $H_1(s, d)=0$. We can rephrase this condition by 
saying that (\ref{exact}) will always be true if 
the anomaly consistency condition $s A = d B$ has only 
trivial solutions $A=s A_1 + d B_1$. 

The general case goes along similar lines. 
Consider the case where the algebra 
of the symmetry transformation is given by 
\be \label{alg}
[s(\e), s(\h)]\f^A = s(\e \xx \h) \f^A
\ee
(for simplicity, we consider closed symmetry algebra)
where $s(\e) \f^A = (s \f^A)^a \e_a$, $\e_a$ and $\h_a$ are the parameters 
of the symmetry transformations, $a$ is an algebra index,  
$(\e \xx \h)^a = f^a{}_{bc} \e^b \h^c$, and $f^a{}_{bc}$ are the structure
constants of the algebra. 
 
The Wess-Zumino consistency condition for the integrated anomaly reads\cite{WZ}
\be \label{WZconsist}
A_{int}(\e \xx \h) = s(\e) A_{int}(\h) - s(\h) A_{int}(\e)
\ee
where $A_{int}(\e)=\int A(\e) = \int A^a \e_a$.
One may easily check that 
\be \label{trivial}
A(\e)=s(\e) K + \pa_\m A^\m
\ee
is a solution of (\ref{WZconsist}) for any $K$. In the case 
of nilpotent symmetries, $f^a{}_{bc}=0$, equation (\ref{WZconsist})
implies that the anomaly is $s$-closed. Furthermore, the trivial solution
(\ref{trivial}) correspond to an exact solution.
  
Assuming that (\ref{WZconsist}) has only the (trivial) solution (\ref{trivial})
we now show that the theory is stable. Consider (\ref{stab}).
Let us first make explicit in our notation the parameter of the transformation
and suppress the $\hbar^M$ as our considerations hold at any order in $\hbar$,
\be \label{stab1}
s(\h) \cl + L_1(\h) = \pa_\m C^\m (\h)
\ee
Act in this equation with $s(\e)$ and antisymmetrize in $\e$ and $\h$.
Using (\ref{alg}) we get
\be
s(\e \xx \h) \cl + s(\e) L_1(\h) - s(\h) L_1(\e) = 
\pa_\m (s(\e) C^\m(\h) -s(\h) C^\m(\e))
\ee
Using again (\ref{stab1}) to eliminate $s(\e \xx \h) \cl$
we obtain a local form of (\ref{WZconsist}).
Therefore, since by assumption this equation has only the 
trivial solution (\ref{trivial}) we obtain
\be
L_1 = s M + \pa_\m C^\m
\ee
which just (\ref{exact}) with $M^\m \to C^\m$.
This finishes the proof that the theory is stable if the anomaly consistency
condition has only trivial solutions.

\subsection{Invariance of the $S$-matrix}

In this subsection we analyze point 4, namely the question of equivalence 
between  conditions (\ref{cons}) and (\ref{qn}).
Having established this equivalence the invariance of the $S$-matrix 
follows as discussed in section 4.1.

At order $g$ the equivalence has been already established 
at (\ref{eq1}). There we have explicitly shown that both 
general symmetry conditions,
(\ref{cons}) and (\ref{qn}), pose the same condition on the
physical coupling $T_1$, namely 
\be
\label{qn11}
[Q, T_1 \} = \pa_\m T^\m_{1/1},  
\ee
where $ T^\m_{1/1} = \cl_1^\m $.
We now define 
\be 
\label{tnl}
T^\m_{n/l}=T_n [T_1(x_1) \cdots T^\m_{1/1}(x_l) \cdots T_1(x_n)].
\ee
Let us also use the notation
\be
\pa_\m  \ct^\m_n(x_1, \ldots, x_n) = \sum_{l=1}^m \pa_\m^l T^\m_{n/l}.
\ee

We shall now show that the condition (\ref{qn}), namely 
$ [ Q , T_n \} = \pa_\m  \ct^\m_n $,
with this definition
implies the same conditions on the 
the time-ordered products $T_n [T_1... T_1]$
as the Quantum Noether condition (\ref{cons}), given by 
$\pa_\m \cj_n^{\m} = 
\sum_{l=1}^n \pa_\m^l \cj_{n/l}^{\m} = 0 $ 
where 
$\cj^\m_{n/l}=T[T_1 (x_1) \cdots j_0^\m(x_l) \cdots T_1(x_n)]$.
In this sense the two general symmetry conditions
are called equivalent.

Because Poincar\'{e} invariance and causality already 
fix the time-ordered products $T_n [T_1... T_1]$
up to the local normalization ambiguity $N_n$, we only have to 
show that these local normalization terms  $N_n$ are constrained 
in the same way by both conditions, (\ref{cons}) and (\ref{qn}). 

The inductive proof of condition (\ref{qn}) proceeds along the same
lines as the one of (\ref{cons}) (as explained around
formula (\ref{cons2})): Assuming that the condition (\ref{qn}) 
is satisfied for all $m<n$ we can directly derive the fact 
that the condition 
at the $n$th order can only be violated by a local distribution
(for a detailed proof see \cite{H95}, section 2b):
\be
\label{qnmod}
[ Q, T_n ] =\pa_\m  \ct^\m_n + A_n
\ee
We shall now discuss in some detail the local terms $A_n$ 
arising from tree level graphs:
To find the local tree-level obstruction terms $A_n$ 
that arise in the right hand side of (\ref{qnmod}) we use the 
same off-shell procedure as before. Namely, we first differentiate, 
keep only the  field equation terms and then do the contractions.
Thus, to get the local terms at $n=2$ we first rewrite (\ref{qn11}) 
including also 
the off-shell terms:  
\be
\label{qn111}
\pa_\m T^\m_{1/1},
- [Q, T_1 \} =
-  s_1 \f^A \ck_{AB} \f^B 
\ee
Then we get at the next order - using the natural splitting solution
(denoted by the subscript `c')- :
\be
\pa_\m \ct^\m_{c, 2}(x_1, x_2) -
[Q, T_{c,2}(x_1,x_2)] =
-  (\hbar/i)  (2 s_1 T_1 - \pa_\m (\frac{\pa T_1}{\pa(\pa_\m \f^A)} s_1 \f^A)) 
\d(x_1,x_2)
\label{222}
\ee  
Now we add the local normalization ambiguity $N_2$, which is uniquely defined 
with respect to the natural splitting solution $T_{c,2}$, to the equation, 
\be
T_2[T_1(x_1) T_1(x_2)] = T_{c,2}[T_1(x_1) T_1(x_2)] + N_2 \d(x_1,x_2) 
\ee
We get 
\bea
\pa_\m \ct^\m_{c, 2}(x_1, x_2) - [Q, T_2(x_1,x_2)] &=&
- (\hbar/i) [2! (s_1 T_1 + s_0 ({1 \over 2} N_2)) \nonu
&&- \pa_\m (\frac{\pa T_1}{\pa(\pa_\m \f^A)} s_1 \f^A)] \d(x_1,x_2)
\label{333}
\eea

Using the same identification
$T_1=(i/\hbar)\cl_1$, $N_2/2!=(i/\hbar)\cl_2$ as before 
and comparing (\ref{333}) with 
the corresponding formula of the Quantum Noether condition
(\ref{n=3}), we see that we ``miss'' the 
term $(\pa \cl_2/\pa(\pa_\m \f^A))s_0 \f^A$. This term 
``covariantizes'' the current $j_0^\m$ that generates the linear 
transformation $s_0 \f^A$. In the present case,  
we have started with the $T_{1/1}^\m$ coupling instead of the 
$j_0^\m$ coupling, so we do not expect to find these 
``covariantization" terms. In addition, what matters is how 
the normalization freedom of the correlation functions without 
a current insertion are fixed. 
Also, now the combinatorial factor is $2!$ instead of 
$3!$. This is due to the fact that we are in 2nd order instead 
of 3rd. From (\ref{333}) we see that  condition (\ref{qn}) at order
$n=2$ is satisfied if and only if
\be \label{noether3new}
s_1 \cl_1 + s_0 \cl_2 = \pa_\m \cl_2^\m + s_2 \f^A \ck_{AB} \f^B
\ee
This coincides with the condition on the normalization terms,
$\cl_1, \cl_2$,  we derived from the Quantum Noether condition
(see \ref{noether3}).

We now fix the tree-level normalization freedom of 
$T_2 [T_{1/1}^\m T_1]$
\be
\label{tyty}
T_2 [T_{1/1}^\m(x_1) T_1(x_2)] = T_{c,2} [T_{1/1}^\m(x_1) T_1(x_2)]
-  j_2^\m \d(x_1, x_2)
\ee
by  condition (\ref{qn}). Then we get
\be
j_2^\m=2!(- \cl_2^\m)
+ \frac{\pa \cl_1}{\pa(\pa_\m \f^A)} s_1 \f^A. 
\ee
So we end up with
\be \label{ct2}
\pa_\m \ct^\m_{2}(x_1, x_2) - [Q, T_2(x_1,x_2) = 
-  2! (s_2 \f^A) \ck_{AB} \f^B
\ee
After imposing the free field equations, we arrive at  condition 
(\ref{qn}). 

Following exactly the same techniques as the ones presented
 in subsection 4.3  
one shows that condition (\ref{qn})
fixes all tree-normalization terms in the same way as our
Quantum Noether condition (\ref{cons}). The derivation of the
Noether consistency equations is totally analogous up to different
combinatorial factors and up to covariantization terms.
For the purpose of completeness, we state here the formulae
analogous to the equations  we got from the 
Quantum Noether condition.
In the general case, for $ 1 \leq n \leq k $  (k is defined as 
in section 4.3, namely by the condition $s_m=0$ for all $m>k$),  
the result for the tree-level normalization freedom of 
$T_n [T_{1/1}^\m T_1 \dots T_1]$ (with respect to the 
natural splitting solution) is 
\be \label{jnnew}
j_{n}^\m= -n!\cl_{n}^\m
+(n-1)! \sum_{l=1}^{n-1} l \frac{\pa \cl_{n-l}}
{\pa(\pa_\m \f^A)} s_l \f^A 
\ee
where we have used the definition in (\ref{lagdef}).
To derive this result one may use again the  distributional identity
(\ref{ident}). In addition, we get 
\be \label{ctnew}
\pa_\m \ct^\m_{n} - [Q, T_n ] = 
- n! (s_n \f^A) \ck_{AB} \f^B
\ee
For the general case $n > k$ one gets 
\bea
\label{highernew}
\pa_\m \ct^\m_{n} -  [Q, T_n ] &=&
- n![s_0 \cl_{n} + s_1 \cl_{n-1}
+ \cdots + s_k \cl_{n-k} ] \nonu
&&+(n-1)! \pa^\m \sum_{l=1}^{k} 
l \frac{\pa \cl_{n-l}}{\pa(\pa_\m \f^A)} s_{l} \f^A
\eea
where again use of (\ref{ident}) has been made. 
This equation implies that 
\be \label{n>knew}
s_0 \cl_{n} + s_1 \cl_{n-1} 
+ \cdots + s_k \cl_{n-k}=
\pa_\m \cl_{n}^\m.
\ee
which coincides exactly with the constraint on the local
normalization terms, $\cl_{n}$, we got from the Quantum Noether condition 
(see (\ref{n>k})).
In full analogy with the analysis there,  
one achieves 
\be 
\label{hhh}
\pa_\m \ct^\m_{n} -  [Q, T_n ]        = 0
\ee
by renormalizing the current as
\be \label{jn1new}
j_{n}^\m=-n!\cl^\m_{n} + (n-1)! \sum_{l=1}^{k} 
l \frac{\pa \cl_{n-l}}{\pa(\pa_\m \f^A)} s_{l} \f^A
\ee
and without the need to use the free field equations, 
which corresponds to the fact that there is no 
field-equation term in (\ref{hhh}).
 
Comparing  formulae (\ref{jnnew}),(\ref{jn1new}) with 
formulae (\ref{jn}),(\ref{jn1}), we conclude that 
the (renormalized) Noether currents derived from the 
two different symmetry conditions, (\ref{qn}) and (\ref{cons}),
also coincide up to different
combinatorial factors and up to covariantization terms.
In addition, the Lagrangian $\cl$ constructed out of the 
tree-level normalization terms (that follow from  
condition (\ref{qn})) is invariant (up to total derivatives)
under the symmetry  transformation,
\be
s \f^A = \sum_{l=0}^k g^l s_l \f^A.
\ee

The issue of stability can be analyzed in exactly the same way as in 
section 4.4. One shows that  condition (\ref{qn}) at loop level
implies that the normalization ambiguity at the loop level, $N_n(\hbar)$,
is constrained in the same way as the tree-level normalizations, 
$N_n(\hbar^0)$. Once the stability has been established
the equivalence of (\ref{cons}) and (\ref{qn}) at loop level follows.
Condition (\ref{qn}) guarantees the 
invariance of the $S$-matrix
under the corresponding asymptotic symmetry in the adiabatic limit. 
In the case of BRST symmetries
the asymptotic linear part of the symmetry  directly 
implies the unitarity of the physical S-matrix, i.e. 
the crucial decoupling of the unphysical degrees of freedom (\cite{Kogu},
see also \cite{H95}, chapter 7).
So the Quantum Noether condition (\ref{cons}) also directly implies this 
crucial property in the case of BRST symmetries.

\section{Examples}
\setcounter{equation}{0}

\subsection{Yang-Mills Theory}
In this section we present the construction
of the Yang-Mills theory with gauge group $G$.  According to the discussion 
in section 3, we first need to specify a set of free field and to give 
their commutation relations. For the YM theory, one has 
the YM field $A_\m^a$, the ghost $c^a$ and the anti-ghost $b^a$,
where $a$ is an gauge group index. We shall discuss the theory 
in the Feynman gauge. For  other gauge choices we refer to 
\cite{H97,Aste98}. The fields satisfy the 
free-field equations
\be \label{ymfeq}
\Box A_\m^a=0; \ \ \Box c^a=0; \ \ \Box b_a=0,
\ee
and the (anti)-commutation relations,
\bea \label{comYM}
[A_\m^{(-) a}(x), A_\n^{(+) b}(y)]&=&
i \hbar \d^{ab} g_{\m \n} D^{+}(x-y) \nonu
\{b^{(-)}_a(x), c^{(+)}{}^b(y)\}&=&-i \hbar \d_{a}^{b} D^{+}(x-y)
\eea
where the super-index $\pm$ designates the emission and absorption 
parts of the corresponding field
and $D^{\pm}$ is the (zero mass) Pauli-Jordan distribution (see appendix A). 
     
The field equations in (\ref{ymfeq}) may be derived from the 
following free Lagrangian 
\be \label{YMac}
\cl_0 = - \frac{1}{4} F^{\m \n} F_{\m \n} + b \Box c 
- {1 \over 2} (\pa{\cdot}A)^2 
\ee
where $F_{\m \n}{=}\pa_\m A_\n -\pa_\n A_\m$. 
With these conventions, the kinetic operator 
$\ck_{AB} \f^B= 
\pa_\m(\frac{\pa \cl_0}{\pa (\pa_\m \f^A)})-{\pa \cl_0 \over \pa \f^A}$, 
and the corresponding propagator $\cd$ are 
equal to $\ck_{AA} A=-\Box A, \ck_{bc} c=-\Box c, \ck_{cb} b=\Box b$,  
and, $\cd^{AA}=i \hbar g_{\m \n} \d_{a b} D_F(x-y)$ and 
$\cd^{bc} = -i \hbar \d_{a}^b D_F(x-y)$, where 
$\Box D_F(x-y) = \d(x-y)$ (see appendix A).

The free Lagrangian is invariant under the following BRST transformations
\be \label{var}
s_0 A_\m^a = \pa_\m c^a \L; \ \  s_0 c^a =0; \ \ 
s_0 b^a= - \pa \cdot A^a \L   
\ee
where we have introduced an anticommuting constant $\L$. 
With this parameter present $s_0$ is a derivation\footnote{One  
may prefer to use an anti-derivation $\s_0$ instead 
of the derivation $s_0$, 
since the current is fermionic.
In this case the parameter $\L$ will not be present in (\ref{var}).
To get the same signs in both cases one should use the Leibniz 
rule $\s_0 (AB) = A \s_0 B + (-1)^B (\s_0 A) B$, where $(-1)^B$ 
denotes the grading of $B$. (We use the convention that $\L$ is 
removed from the right side of the equations. This choice is 
co-related with the convention chosen in (\ref{GC}) to have the test 
functions to the right of the $T$-products. See the discussion 
in the paragraph after (\ref{qc})).  
We choose $s_0$ in the following in order to show how 
the general analysis presented in section 4 (which involves a bosonic current)
is also applicable to examples where the Noether current
is fermionic.}.  One may derive the Noether current by making 
$\L$ local, varying the Lagrangian and collecting all terms that are 
proportional to $\pa_\m \L$ as explained in section 2. 
The result is 
\be
j_0^\m = - \pa_\n c F^{\m \n} - (\pa \cdot A) \pa^\m c.
\ee
(the overall sign is fixed by our convention to always remove the 
anticommuting variable from the right side of the equations).
The current is conserved when the free field equations are satisfied.  
In particular,
\be \label{s0}
\pa_\m j_0^\m \equiv \pa_\m \cj_1^\m
= \pa_\m c (-\Box A^\m) + (\pa \cdot A) (-\Box c).
\ee
This corresponds to  the general formula (\ref{n1}) in section 4:
\be
\pa_\m j_0^\m \equiv  \pa_\m \cj_1^\m = (s_0 \f^A) \ck_{AB} \f^B.
\ee

The corresponding BRST charge is given by
\be
Q=\int j_0^0 d^3x = - 
\int (\pa \cdot A^a) \stackrel{\leftrightarrow}{\pa^0} c^a d^3x
\ee
(with the above stated conventions, 
$s_0 \f^A = (i/\hbar) \{Q , \f^A] \L$).

One may easily check that $s_0$ is nilpotent ($s_0 (\L_1) s_0(\L_2)=0$) 
when the free-field equation are satisfied. This is particular to the 
case of BRST symmetry\footnote{Let us mention that in the 
Lagrangian approach one may incorporate global symmetries into the 
BRST operator by introducing constant ghost fields (see, for 
instance, \cite{BHW}). In the EG formalism, where one starts with 
free fields in the asymptotic Fock space, there is no 
natural way to incorporate constant ghost fields. 
We shall, therefore, not discuss further this possibility.}.     
In addition, one has the ghost charge (which can also be obtained
as a Noether current),
\be \label{qc}
Q_c = {i \over \hbar} \int d^3 x b_a \stackrel{\leftrightarrow}{\pa^0} c^a.
\ee
The ghost charge introduces a grading, the ghost number,
in the algebra generated by the fundamental field operators.
The ghost number of the gauge field $A_\m^a$ is zero, of 
ghost field $c^a$ is $+1$, and of the antighost field $b_a$ is
$-1$. It follows that the BRST charge has ghost number $+1$. 

To obtain the theory in the Epstein-Glaser approach one starts 
with the coupling $g_\m j_0^\m$ and a yet unknown coupling $g T_1$.
Since $j_0^\m$ is fermionic, $g_\m$ is fermionic too, 
and one has to be careful with signs.  
A practical trick that helps keeping track of them is to write 
$g_\m=\L g'_\m$,
where $\L$ is an anticommuting constant (as in (\ref{var})). 
Now, let us define $j_0'{}^\m= j_0^\m \L$.
Since $j_0^\m g_\m = j_0'{}^\m g'_\m$ the $S$-matrix constructed 
with a coupling  involving the bosonic current $j_0'{}^\m$ smeared out 
by the bosonic test function $g'_\m$ is the same with the one constructed
with a coupling involving the anticommuting current $j_0^\m$
smeared out by the fermionic test function $g_\m$.
However, since all vertices are now bosonic one need not 
worry about signs. At the end we are interested in the $T$-products 
that involve the fermionic current. To obtain those, one simply 
pushes $\L$ to the right where it recombines with $g'_\m$ to give $g_\m$. 
This automatically produces all correct signs. If 
one considers multi-current correlation functions one introduces as many 
anti-commuting constants as the number of current insertions, so one may 
need infinite number of anti-commuting constants.  

As we have seen the Quantum Noether condition 
(\ref{cons}) at second order, $\pa_\m \cj_2^\m(x_1, x_2)=0$,
is equivalent to the condition
\be \label{lev1}
s_0 \cl_1 = \pa_\m \cl_1^\m   
\ee
where $\cl_1=(\hbar /i) T_1$.
Observe that both sides of this equation involve a nilpotent differential;
the left hand side the (abelian) BRST differential and the 
right hand side the co-differential 
$\d=*d*$, where $*$ is the Hodge operator and $d$ is the exterior derivative.
(Of course, the construction does not depend on the metric). Therefore, 
$\cl_1 \in H^{weak}_0(s_0, d)$ (the sub-index $0$ denotes ghost number
and denomination ``weak'' that we are working modulo field equations).
Notice, however, that $d$ is not
acyclic since we are working on a space where the free field equations 
are satisfied. The latter introduce non-trivial cycles. An example 
of the latter at ghost number $2$ is $C_\m=c^a \pa_\m c^a$. One may check 
that $\pa^\m C_\m=0$, but $C_\m \neq \pa_\m B$ for any $B$. 

The most general solution of (\ref{lev1}) is equal to  
\footnote{We note that the  well-known Curci-Ferrari mass term 
\cite{Ferrari} $ \cl'_1 = m^2 (1/2 A_\mu^a A_a^\mu + c_a b_a) $ 
is compatible with (\ref{lev1}). 
Had we included this term in (\ref{l1}) at order $g$ the Quantum Noether 
method would not be fulfilled at the next order as explicitly
shown in \cite{H97}, section 3. If one 
sums up all diagrams with mass insertions as one regards the mass vertex 
to be of order $g^0$ (this is the same as starting  
with free massive gauge fields), then the Quantum Noether method  yields 
the  Curci-Ferrari model\cite{Ferrari}. However, in this case the final 
transformations are not nilpotent any longer. One may still prove that 
the theory is renormalizable, but it is not unitary\cite{Cu-Fe}.
Including an unphysical scalar 
(St\"ueckelberg model) nilpotency can be restored but this theory 
can be shown to be only `generalized renormalizable' (see also 
\cite{dragon2,H96}). If one adds a 
physical scalar (Higgs field) to the theory one ends up with a unitary 
and renormalizable theory with massive gauge bosons (for a discussion
in the EG framework, see \cite{Aste97}).} 
\be \label{l1}
\cl_1 = g f^{abc} ({1 \over 2} A_{\m a} A_{\n b} F^{\n \m}_c
+ A_{\m a} c_b \pa^\m b_c) + s_0 C + d D.
\ee
where $C$ and $D$ are known local terms. We shall show in \cite{paper2} that 
the exact terms do not change the physics. In the following we 
concentrate on the non-trivial terms. We briefly discuss  
the exact terms afterwards.  

The Quantum Noether condition at order $g$ implies that the structure 
constants $f_{abc}$ are totally antisymmetric, 
(see for instance \cite{antiN}, \cite{Stora2}), 
but still not further restricted by a Jacobi identity. This contraint
follows from the symmetry condition at order $g^2$ (see below).
Let us analyze further (\ref{cons}) at second order. 
We shall present this calculation in some detail 
in order to illustrate how one deals with the various subtle points
discussed in section 4. 
We are interested in computing 
\be
\pa_\m \cj_{c, 2}^\m(x_1, x_2)= 
\pa_{x_1}^\m T_{c,2}[j_\m^0(x_1) T_1(x_2)] + 
\pa_{x_2}^\m T_{c,2}[T_1(x_1) 
j_\m^0(x_2)]
\ee
at tree level. Let us start by first computing 
$\pa_{x_1}^\m T_{c,2}[j_\m^0(x_1) T_1(x_2)]$. 
Since the tree-level Wick contractions
satisfy the Leibniz rule, and $T_1$ only depends only on the 
fields and their first derivative, one can first compute
$\pa_{x_1}^\m T_{c,2}[j_\m^0(x_1) \f^A(x_2)]$ and  
$\pa_{x_1}^\m T_{c,2}[j_\m^0(x_1) \pa_\m \f^A(x_2)]$, 
where $\f^A=A_\m^a, c^a, b^a$,
and then use the Leibniz rule. To correctly take care  
of the various signs
we insert an anti-commuting constant $\L$ next to $j_0^\m$.
As has been argued in detail in section 4, one can first move the derivative
inside the $T$-product and then do the contractions. Let us compute
$\pa_{x_1}^\m T_{c,2}[j_\m^0(x_1)\L b^a(x_2)]$
After the first step 
and before the contractions one has (using (\ref{s0}))
\be
[\pa_\m c (-\Box A^\m) + (\pa \cdot A) (-\Box c)](x_1) \L b_a (x_2)
\ee
Clearly, a local term is produced when there is a contraction between
the second term in the square brackets and $b_a(x_2)$. 
It is equal to $(\hbar/i) (- \pa^\m A^a_\m \L)$, which up to $(\hbar/i)$
(which is there to cancel the overall $(i/\hbar)$), is equal to $s_0 b_a$
as it should. 
One can also contract the first term with $b_a(x_2)$. This yields 
a non-local 
term proportional to the $A$ field equation. 
This is an example of ``irrelevant contraction''. As explained in section 4
these terms are irrelevant and they will be discarded. 
We shall, from now on, only concentrate 
on the ``relevant'' contractions, namely the ones resulting from 
contractions involving the field $\f^B$ in $\ck_{AB}\f^B$.

Following the procedure we have just outlined one obtains,
\bea
\pa_\m \cj_{c, 2}^\m(x, y)
&=& g f_{abc} [ 2 \pa_\m c^a A_{\n}^b F^{\n \m c} + 
2 \pa_\m c^a c^b \pa^\m b^c - 2 A_{\m}^a c^b \pa^\m \pa{\cdot}A^c \nonu
&&+\pa^\m (A_\m^a c_b \pa{\cdot}A^c + A_\m^a A_\n^b \pa^\n c^c)]\L 
\d(x_1, x_2) \nonu
&=& \{\pa^\m [2(s_1 A^{\n}_a F^a_{\m \n} - s_1 c^a \pa_\m b_a)
+g f_{abc}(A_\m^a A_\n^b \pa^\n c^c + A_\m^a c^b \pa^\n A_\n^c) \L] \nonu
&&+2(s_1 A^{\m a} (-\Box A_\m^a) + s_1 c^a \Box b^a)\} \d(x-y) 
\label{n2ym}
\eea
where 
\be
s_1 A^a_\m = g f^a{}_{bc} A_\m^b c^c \L; \ \ 
s_1 c^a = {g \over 2} f^a{}_{bc} c^b c^c \L
\ee
Notice that the field equation terms in (\ref{n2ym}) were created in the 
process of factoring out a total derivative from the rest of the terms.
Let us now fix the tree-level ambiguity such that (\ref{cons}) at second order
holds when the free-field equations are satisfied.
To this end, we let
\be
T[j_0^\m(x_1) T_1(x_2)] = T_c[j_0^\m(x_1) T_1(x_2)] + j_1^\m \d(x_1-x_2)
\ee
Then (\ref{cons}) implies
\be
j_\m^1 = -2 g f_{abc}(A^{\n b} c^c F^a_{\m \n} 
+ {1 \over 2} c^b c^c \pa_\m b^a)
-g f_{abc}(A_\m^a A_\n^b \pa^\n c^c + A_\m^a c^b \pa^\n A_\n^c)
\ee
The first two terms in the Noether current are the ones that 
generate the $s_1$ transformation. The last two  ``covariantize'' 
$j_0^\m$. All of them are part of the Noether current of the 
non-linear theory up to combinatorial factors 
which take care of the additional factors in the perturbative expansion,
\be \label{jna}
j_\m({\rm non{-}abelian}) = -D_\n c^a \cf_{\m \n}^a
- (\pa\cdot A^a) D_\m c^a - {1 \over 2} g f_{abc} c^a c^b \pa_\m b^a,
\ee
where 
\be
\cf_{\m \n}^a = \pa_\m A_\n^a - \pa_\n A_\m^a + 
g f^a{}_{bc} A_\m^b A_\n^c; \quad
D_\m c^a = \pa_\m c^a + g f^a{}_{bc} A_\m^b c^c.
\ee
So, finally at $n=2$ we have off-shell,
\be \label{offshell2}
\pa_\m \cj_2^\m (x, y) = 
2[s_1 A^{\m a}(-\Box A_\m^a) + s_1 c^a \Box b^a]\d(x_1, x_2)) 
\ee
This corresponds to the general formula 
\be
\pa_\m \cj_2^\m (x_1, x_2) = 2! s_1 \f^A \ck_{AB} \f^B.
\ee

One may also check that $\cl_0 + \cl_1$ is equal to the YM action,
\be \label{YM}
\cl_{YM}=  -{1 \over 4} \cf^{\m \n} \cf_{\m \n} + b \pa^\m D_\m c
-{1 \over 2} (\pa \cdot A)^2,
\ee
but the four-gluon term. The latter is of order $g^2$ and will be 
recovered at next order.
 
To examine (\ref{cons}) at third order we need the second-order result 
off-shell (\ref{offshell2}).
Using 
\be 
\pa_\m \cj^{\m}_{c, 3}(x_1, x_2, x_3) =
\pa_\m \cj^{\m}_{2}(x_1,x_2) T_1(x_3)
+\pa_\m \cj_1^\m(x_1) N_2(x_2, x_3)
+ {\rm cyclic\ in}\ x_1, x_2, x_3, 
\ee 
one gets at $n=3$,
\bea 
\pa_\m \cj^{\m}_{c, 3}(x_1, x_2, x_3)&=& 
[3! (g^2 f_{abe} f_{ecd} A_\m^a A_\n^b A_\m^c \pa_\n c^d
 + s_0 \cl_2)\L \nonu
&-& 2! \pa_\m (-2 g f_{abc} A_\m^a A_\n^b s_1 A_\n^c
+\frac{\pa \cl_2}{\pa(\pa_\m \f^A)}s_0 \f^A)] \d(x_1, x_2, x_3)
\eea
where $\cl_2 = (\hbar/i) N_2/2!$ and $N_2$ denotes the unique local 
normalization term of $T_2$ with respect to $T_{c,2}$. From here 
we determine $\cl_2$, which is  just the missing
four-gluon coupling in (\ref{YM}), and also the Noether current 
renormalizations,
\bea
&&\cl_2 = - {1 \over 4} (g f_{abc} A_\m^b A_\n^c)^2, \nonu
&&j_2^\m = -4 g^2 f_{abc} f_{cde} A_\m^a A_\n^b A_\n^d c^e \label{l2}
\eea
where $j_2$ is defined as in (\ref{j2def}). 
Notice that this is precisely the ``covariantization'' term missing from 
(\ref{jna}). We note that one also finds
the Jacobi
identities for $f_{abc}$ as a consequence of the Quantum Noether
Condition at order $n=2$. At $n=3$ we finally have even off-shell
\be \label{n3ym}
\pa_\m \cj^{\m}_{3}(x_1, x_2, x_3)=0
\ee
Notice the absence of field equation terms in the right hand 
side of (\ref{n3ym}).
This means that no new tree-level local terms will emerge at higher 
orders.  

One may easily check that $s_0$ and $s_1$ as defined above satisfy
$\{s_i, s_j\}=0$, where $i=0, 1$. It follows that $s=s_0+s_1$
squares to zero. It has been argued that only non-trivial solutions
of (\ref{lev1}) are physically relevant. This question will be analyzed in 
detail in \cite{paper2}. Let us already briefly discuss this issue 
here. Suppose that instead of
$\cl_1$ in (\ref{l1}) one considers $\cl_1 + s_0 C$. This is still 
a solution of (\ref{lev1}). At next order, however, one obtains
the equation
\be
s_1 (s_0 C) + s_0 \cl_2'=0
\ee
where $N_2=(i/\hbar)(\cl_2 + \cl_2')$ and $\cl_2$ is as in (\ref{l2}).
It follows that $\cl_2'=s_1 C$. Thus, adding an exact $s_0$ term in 
$\cl_1$ results in the addition of an $s$-exact term in the Lagrangian.
The statement that such $s$-exact term are physically irrelevant
will be worked out in the EG formalism in \cite{paper2}. 
In our specific example one may check 
that adding the exact terms, $\cl_1'= \beta_1 s_0 (f_{abc} \pa A^a c^b c^c)$
and $\cl_1''= \beta_2 \pa^\mu ( f_{abc} c^a b^b A_\mu^c )$ with the free
constants $\beta_1$ and $\beta_2$, 
results in additional terms in the  transformation rules at order
$g$ and in the well-known four-ghost coupling at order $g^2$.

Let us now move to loop level. It is well known\cite{Dixon} that in order
to have a candidate anomaly (i.e. non-trivial element of $H^1(s, d)$). 
one needs a non-vanishing $d_{abc}$ and an epsilon symbol $\e_{\m\n\r\s}$.
However, in a theory without
chiral fermions one does not have an epsilon symbol. So, in
pure Yang-Mills the non-trivial elements of $H^1(s, d)=0$ will not occur.
According to the analysis presented in section 4.4  
this is sufficient to guarantee that the loop normalization ambiguity 
is constraint the same way as the tree-level one.
We therefore conclude that the Yang-Mills theory is renormalizable.  

\subsection{The $N=1$ Wess-Zumino Model}

We now turn to our supersymmetric example.
The field in Wess-Zumino model\cite{WZ1} are a complex scalar field
$\f$ and its fermionic partner $\psi^\a$.
We use the two component spinor notation of \cite{superspace} 
(see appendix A).
The fields satisfy the following field equations 
\be
\Box \f=0; \ \ \pa_{\a \dda} \j^\a=0. 
\ee
The commutations relations are 
\bea
&&[\f^{(-)}(x_1), \bar{\f}^{(+)}(x_2)] = i\hbar 2 D^{+}(x_1,x_2) \nonu
&&\{ \psi_\a{}^{(-)}(x_1), \psi_{\dda}{}^{(+)}(x_2) \}= i\hbar 
S^+_{\a \dda}(x_1,x_2)
\eea 
where $S_{\a \dda}^+= 2 i \pa^{\a \dda} D^{+}$.

The field equations can be derived from the Lagrangian
\be
\cl_0 = {1 \over 2} \bar{\f} \Box \f + \j^{\dda} i \pa^\a{}_{\dda} \j_\a
\ee
With these conventions $\ck_{\bar{\f}\f} \f = - \Box \f/2$,
$\ck_{\f \bar{\f}} \bar{\f} = - \Box \bar{\f}/2$,
$\ck_{\dda \a} \j^\a = -i \pa^\a{}_{\dda} \j_\a$ and 
$\ck_{\a \dda} \j^{\dda} = -i \pa_\a{}^{\dda} \j_{\dda}$. 
The corresponding
Feynman propagators are given by $\cd^{\f \bar{\f}}=2 i \hbar \D_F(x-y)$,
and $\cd^{\a \dda}=i\hbar S_F^{\a \dda}$, where 
$S_F^{\a \dda}= 2 i \pa^{\a \dda}_x D_F(x-y)$. For a derivation of these 
formulae   see appendix A. 

This action is invariant under 
the linear supersymmetry transformations,
\be
s_0 \f = - \e^\a \j_\a, \ \ s_0 \bar{\f} = - \e^{\dda} \j_{\dda}, \ \ 
s_0 \j^\a = - \e^{\dda} i \pa^\a{}_{\dda} \f, \ \
s_0 \j^{\dda} = - \e^{\a} i \pa_\a{}^{\dda} \f
\ee
The associated Noether current is equal to 
\be
j_0^{\a \dda}=\e^\b \j^\a \pa_\b{}^{\dda} \bar{\f}
+\e^{\ddb} \j^{\dda} \pa^\a{}_{\ddb} \f
\ee
An easy calculation yields
\bea
\pa_{\a \dda} j_0^{\a \dda} &=& 
(- \e^\a \j_\a)(-\half \Box \bar{\f}) 
+ (- \e^{\dda} \j_{\dda})(-\half \Box \f) \nonu
&&+(- \e^{\ddb} i \pa^\a{}_{\ddb} \f) (-i\pa_\a{}^{\dda} \j_{\dda})
+(- \e^{\b} i \pa_\b{}^{\dda} \f) (-i\pa^\a{}_{\dda} \j_\a)
\eea
In a similar way as in the Yang-Mills example, one may check that 
this current correctly produces the supersymmetry variation 
inside correlation functions.

Invariance at first order requires that we find a $T_1=(i/\hbar) \cl_1$ such 
that 
\be 
s_0 \cl_1 = \pa_\m \cl_1^\m   
\ee 
holds, up to free field equations, for some $\cl_1$ and $\cl_1^\m$. 
The latter are constraint by power counting. 
The most general solution (for simplicity we restrict ourselves 
to a massless theory) is
\be
\cl_1= (1/2) (\f \j^2 + \bar{\f} \bar{\j}^2)
\ee
where $\j^2=\j^\a \j_\a$ and $\bar{\j}^2=\j^{\dda} \j_{\dda}$.
Then,
\bea
\pa_{\a \dda} \cj^{\a \dda}_{c,2}(x_1, x_2) &=& 
[\pa_{\a \dda} i (\e^{\dda} \j^\a 
\f^2 + \e^\a \j^{\dda} \bar{\f}^2) \nonu 
&&+2(- \half \e^\a \bar{\f}^2) (-i\pa_\a{}^{\dda} \j_{\dda})
+2(- \half \e^{\dda} \f^2) (-i\pa^\a{}_{\dda} \j_\a)]\d(x_1, x_2)
\eea
The first term in the right hand side is removed by fixing the tree-level 
ambiguity 
as
\be
T[j_0^{\a \dda}(x_1) T_1(x_2)] = T_c[j_0^{\a \dda}(x_1) T_1(x_2)] 
+ j_1^{\a \dda} \d(x_1-x_2)
\ee
where 
\be
j_1^{\a \dda} = - i
(\e^{\dda} \j^\a \f^2 + \e^\a \j^{\dda} \bar{\f}^2)
\ee
So, we end up with 
\be
\pa_{\a \dda} \cj^{\a \dda}_{2}(x_1, x_2)=
2! (s_1 \j^\a) \ck_{\a \dda} \j^{\dda}
+ 2! (s_1 \j^{\dda}) \ck_{\dda \a} \j^{\a}
\ee
where the new symmetry variations are given by
\be
s_1 \f=0, \ \ s_1 \j^\a = - {1 \over 2} \e^\a \bar{\f}^2, \ \ 
s_1 \bar{\f}=0, \ \ s_1 \j^{\dda} = - {1 \over 2} \e^{\dda} \f^2
\ee

We mow  move to the next order. 
We have
\be 
\pa_\m \cj^{\m}_{c, 3}(x_1, x_2, x_3) =
\pa_\m \cj^{\m}_{2}(x_1,x_2) T_1(x_3)
+\pa_\m \cj_1^\m(x_1) N_2(x_2, x_3)
+ {\rm cyclic\ in}\ x_1, x_2, x_3, 
\ee 
A straightforward calculation yields
\be 
\pa_\m \cj^{\m}_{c, 3}(x_1, x_2, x_3) =
3![(1/2)(-\e^\a \j_\a) \bar{\f}^2 \f + N_2 
(-\e^{\dda} \j_{\dda}) \bar{\f} \f^2 
+ s_0 \cl_2] \d(x_1, x_2, x_3)
\ee 
where $\cl_2 = (\hbar/i) N_2/2$.
This implies that 
\be
\cl_2 = -{1 \over 4} \f^2 \bar{\f}^2
\ee
Hence, 
\be
\pa_\m \cj^{\m}_{3}(x_1, x_2, x_3) = 0
\ee
Therefore, no new local terms will arise in higher orders.

The sum $\cl = \cl_0 + \cl_1 + \cl_2$ is equal to 
\be
\cl={1 \over 2} \bar{\f} \Box \f + \j^{\dda} i \pa^\a{}_{\dda} \j_\a
+{1 \over 2}  g (\f \j^2 + \bar{\f} \bar{\j}^2) -
{1 \over 4} g^2 \f^2 \bar{\f}^2
\ee
which is indeed the Wess-Zumino Lagrangian. 
The transformation rules which are generated by $s=s_0 + s_1$ are given by,
\bea 
&&s \f = - \e^\a \j_\a, \ \ s \bar{\f} = - \e^{\dda} \j_{\dda}, \nonu
&&s \j^\a = - \e^{\dda} i \pa^\a{}_{\dda} \f - {1 \over 2} g \e^\a \bar{\f}^2 
\ \ 
s_0 \j^{\dda} = - \e^{\a} i \pa_\a{}^{\dda} \f- {1 \over 2} g \e^{\dda} \f^2.
\eea
These are also the correct supersymmetry transformation rules.

Finally, we discuss the issue of stability under quantum corrections
(renormalizability). It has been shown in \cite{susy} that there are 
no anomaly candidates in the $N=1$ Wess-Zumino model. From 
our discussion \footnote{ Strictly speaking one would have to 
extend the considerations of section 4.4 to theories with open 
symmetry algebra in order to be applicable to the present example.}  
in section 4.4 immediately follows that 
the loop normalization ambiguity is constrainted in the same 
way as the tree-level one, i.e. the theory is renormalizable.

\section*{Acknowledgements}
We thank Raymond Stora for discussions and comments 
regarding the Epstein-Glaser framework and for a critical reading 
of the manuscript, Peter van Nieuwenhuizen
for discussions and comments regarding the Noether method, and 
Friedemann Brandt, Walter Troost, and Toine Van Proeyen for discussions. 
TH would like to thank the Institute 
for Theoretical Physics in Stony Brook and 
in Leuven 
and KS the Theory Division at CERN and the MPI in Munich 
where  part of this work was completed
for their kind hospitality. 
TH gratefully acknowledges financial support during his stay
in Stanford by Schweizerischer 
Nationalfonds and by Department of 
Energy under contract number DE-ACO3-73SF00515. 
KS is supported by the European 
Commission HCM program CHBG-CT94-0734 and by the European
Commission TMR programme ERBFMRX-CT96-0045.

\appendix

\section{Conventions}
\renewcommand{\theequation}{A.\arabic{equation}}
\setcounter{equation}{0}

In this appendix we list  in detail our conventions and our various
sign choices. 
Since one of the aims of this article is to make contact 
between the EG and the Lagrangian formalism, it is important to have
compatible conventions on both sides. 

We start from a free massless scalar in four dimensions. 
The Lagrangian is given by
\be \label{scalar}
\cl_0 = \half \f \Box \f
\ee
The overall sign in the Lagrangian is such that the Hamiltonian 
is positive definite. (Our convention for metric is 
$\h_{\m \n}=diag(-1, 1, 1, 1)$ and $x^0$ is the 
time variable $t$. For the space coordinates we use either $x^i$ or $\bfx$). 
The canonical momentum is equal to $p=\dot{\f}$.
Then the equal-time commutation relations (ECR) read,
\bea
&&[\f(t, {\bf x}), \f(t, {\bf y})]=
[\dot{\f}(t, {\bf x}), \dot{\f}(t, {\bf y})]=0 
\nonu
&&[\f(t, {\bf x}), \dot{\f}(t, {\bf y})]=i \hbar \d(\bfx-\bfy) \label{ECR}
\eea
The field equation is 
\be
\Box \f =0
\ee
The commutator of two fields in arbitrary spacetime
points is 
\be
[\f^{(\mp)}(x), \f^{(\pm)}(y)= i \hbar D^{\pm}(x-y)
\ee
where $\f^{(\pm)}(x)$ are the absorption and emission parts of $\f(x)$, 
and 
\be
D^{\pm}(x-y) = \pm (-i) \int {d^3k \over (2 \pi)^3 2 \o} e^{\pm ik(x-y)}
\ee
The Pauli-Jordan distribution is then equal to 
\bea
D(x-y)&=&D^+ + D^-= (-i) \int {d^3k \over (2 \pi)^3 \o} \sin k(x-y) \nonu
&=& (-i) \int {d^4k \over (2\pi)^3} \d(k^2) \sgn(k^0) e^{ik(x-y)}
\eea
The Pauli-Jordan distribution has a causal 
decomposition into retarded and 
advanced part as follows,
\be
D_{ret}(x-y) = \theta(x^0-y^0) D(x-y); \ \ 
D_{adv}(x-y) = -\theta(y^0-x^0) D(x-y). 
\ee
The Feynman propagator  
\be
\cd^{\f\f} \equiv \<0|T(\f(x) \f(y))|0\> = i \hbar D_F(x-y)
\ee
where $\Box_x D_F(x-y)=\d(x-y)$. From the latter equation 
one obtains 
\be
D_F(x-y)=\int {d^4k \over (2\p)^4} {e^{ikx} \over -k^2 + i \e}
\ee
It follows that 
\be 
D_F(x-y) = \theta(x^0-y^0) D^+(x-y) - \theta(y^0-x^0) D^-(x-y).
\ee
It is now easy to verify (as described in section 3) 
that after natural splitting the commutator
of two fields is replaced by the Feynman propagator exactly.

We move to the case of gauge fields.
The Lagrangian in the Lorentz gauge is given by
\be
\cl_0 = - \frac{1}{4} F^{\m \n} F_{\m \n} - {1 \over 2} (\pa{\cdot}A)^2 
\ee
The overall sign in the action is fixed by requiring positivity of the 
energy. The result for the commutator and the propagator are,
\bea
&&[A^{(-)a}_\m(x), A_\n^{(+)b}(y)]=i \hbar \h_{\m \n} \d^{a b} D^+(x-y) \nonu
&&\cd^{AA} \equiv \<0|T(A_\m^a(x) A_\n^b(y))|0\>=
i \hbar \h_{\m \n} \d^{a b} D_F(x-y)
\eea
The indices $a, b$ indices are gauge group indices.

We now move to the ghost sector. We take for Lagrangian
\be \label{ghact}
\cl_0 = \int b_a \Box c^a.
\ee
The field equation are $\Box c^a = 0$ and $\Box b_a=0$. 
$b_a$ is antihermitian 
whereas and $c^a$ is hermitian, so that the 
action is hermitian\footnote{Strictly speaking the ghosts
and the action are pseudo-(anti-)hermitian which indicates
that the actual hermiticity properties are defined in respect 
to a sesquilinear form (indefinite metric) and not in respect of the 
(positive definite) scalar product of the one-particle 
Hilbert space of the ghosts.} 
The anti-commutator of a ghost field with an antighost is
\be
\{b^{(-)}_a(x), c^{(+)b}(y)\}=-i \hbar \d_a^b D^+(x-y)
\ee
The Feynman propagator is
\be
\cd^{bc} \equiv \<0|T(b_a(x) c^b(y))|0\>=-i \hbar \d_a^b D_F(x-y)
\ee

For the case of fermions we use the 
two component notation of \cite{superspace}.
With these conventions one avoids using gamma matrices, and the Fierz
identities become a matter of symmetrizing and antisymmetrizing 
spinor indices.

The universal cover of the Lorentz group in four dimension 
is isomorphic to $SL(2, C)$.
The simplest non-trivial representation of the latter is 
the two component complex Weyl spinor $\j^\a$, $\a=+, -$,
(the $(1/2, 0)$ representation). Its complex conjugate representation 
(the $(0, 1/2)$) is denoted by $\j^{\dda}$. 
Greek letters are reserved for spinor two components indices 
and Roman ones for vector indices. Each vector index is equivalent to
one undotted and one dotted index ($\phi^a = \phi^{\a \dda}$).
Indices are raised and lowered using 
the $sl_2$ invariant antisymmetric two dimensional matrix
$C_{\a \b}$. Since $C_{\a\b}$ is antisymmetric, we have to specify
how exactly we use it to raise and lower indices, and our convention
is the so-called `down-hill' rule from left to right for both
the undotted and the dotted sector. For example,
\be
\j^\a C_{\a \b} = \j_\b; \ \ C^{\a \b} \j_\b = \j^\a; \ \ 
\j^{\dda} C_{\dda \ddb} = \j_{\ddb}; \ 
\ C^{\dda \ddb} \j_{\ddb} = \j^{\dda}.
\ee
In addition, we have the following identity
\be
C_{\a \b} C^{\g \d} = \d_{\a}{}^{\g} \d_{\b}{}^{\d}-
\d_{\b}{}^{\g} \d_{\a}{}^{\d},  \label{C-C}
\ee
{}From (\ref{C-C}) we get
\bea
&\ & C^{\a \b} C_{\a \b} = \d_\a{}^{\a} = 2 \\
&\ & C_{\a \b} \chi_\g - C_{\a \g} \chi_\b  = - C_{\b \g} \chi_\a \\
&\ & \chi_\a \chi_\b = - \frac{1}{2} C_{\a \b} \chi^\g \chi_\g
\eea
The last identity is an example of a Fierz identity. 

The Lagrangian for a massless spinor field is given by
\be
\cl_0=\j^{\dda} i \pa^\a{}_{\dda} \j_\a
\ee
The propagator is
\be
\cd^{\a \dda} \equiv \<0|\j^\a (x) \j^{\dda}(y)|0\> 
= i \hbar S_F^{\a \dda} 
\ee
where 
\be
S_F^{\a \dda} = 2 i \pa^{\a \dda}_x D_F(x-y)
\ee

Finally, the commutation relations of two 
Fermi fields at arbitrary spacetime separation are given by
\be
\{\j_\a^{(-)}(x), \j_{\dda}^{(+)}(y)\}=i \hbar S^{+}_{\a \dda}(x-y)
\ee
where 
\be
S^{+}_{\a \dda}(x-y) = 2 i \pa_{\a \dda}^x D^+(x-y).
\ee

\section{Further Properties of the EG Formalism}

\renewcommand{\theequation}{B.\arabic{equation}}
\setcounter{equation}{0}

In this appendix we highlight some further properties of 
the EG formalism.
In particular we discuss
the infrared problem, the construction of interacting fields, 
the problem of overlapping divergences and a few other issues.
\\
$\bullet$ The distributions $T_n$ in (\ref{GC}) are smeared out by 
tempered test functions
$g \in {\cal S}$. This provides a natural regularization of the
physical infrared problem which arises in massless theories because
the tempered test function cuts off the long-distance part of the
distributions. In the construction of quantum electrodynamics, for
example, the infrared problem is fully separated in the causal
formalism.
Ultimately, one is interested in the physical (so-called adiabatic)
limit $g(x) \rightarrow  g \equiv const.$. 
Epstein and Glaser have proven that this limit exists for  
the vacuum expectation values of the $T_n$ distributions
(in the sense of tempered distributions)
in massive theories if a
suitable normalization is chosen (section 8.2 in \cite{EG}). In this
limit the Green functions
possess all the expected linear properties 
such as causality, Lorentz covariance and the spectral
condition. 
The existence of Green functions in the adiabatic limit
for the case of quantum electrodynamics was shown by Blanchard and
Seneor \cite{Blanchard}. We implement our method before the
adiabatic limit. All equations are understood as distributional ones.
So our analysis is also well-defined in massless theories like
pure Yang-Mills theories where the adiabatic limit
is related to the confinement problem which is not expected to be  
solved in the framework of perturbation theory.\\
$\bullet$ We have argued in the introduction that 
the strength of the EG construction lies in the operator
formalism. However, this strength turns into a weakness of the formalism
when one is interested in non-local details of the theory, for example,
when one tries to translate a simple operator condition into relations
of $C$-number distributions or if one discusses properties of a subgroup
of contributions or even a single type of diagram.\\
$\bullet$ The EG
formalism naturally leads to amputated connected Green functions
and not to one-particle irreducible ones.  This complicates
the discussion of the renormalization group in this framework. \\
$\bullet$ Having constructed the most general $S$-matrix one can
construct interacting field operators (compatible with causality and
Poincar\'{e} invariance) (\cite{EG} section 8, \cite{BKS}).

One starts with an extended first order $S$-matrix
\begin{equation}
S(g,g_1,g_2, \ldots) = \int d^4 x \{T_1(x)g(x) + \Phi_1(x)g_1(x) 
+ \Phi_2(x)g_2(x)+ \ldots\}
\end{equation}
where $\Phi_i$ represent certain Wick monomials like
$(i/\hbar) \varphi$ or $(i/\hbar) :\varphi^3:$.
Following Bogoliubov and Shirkov (\cite{BS}), Epstein and Glaser defined 
the
corresponding interacting fields $\Phi_i^{int}$ as functional
derivatives of the extended S-matrix:
\begin{equation}
\Phi_{i}^{int}(g,x) = 
S^{-1}(g,g_1, \ldots) \frac{\delta S(g,g_1,\ldots)}{ \delta g_i} 
\Big|_{g_i=0} 
\end{equation}
One shows that the perturbation series for the interacting
fields is given by the advanced distributions of the corresponding
expansion of the $S$-matrix, namely
\begin{equation} \label{defint}
\Phi_i^{int}(g,x) = \Phi_i(x) + \sum_{n=1}^\infty \frac{1}{n!} 
\int d^4 x_1 \ldots d^4 x_n {\hbar \over i} 
A_{n+1/n+1} (x_1, \ldots, x_n; x),
\end{equation}
where $A_{n+1/n+1}$ denotes the advanced distributions with $n$
original vertices $T_1$ and one vertex $\Phi_i$ at the $(n+1)$th
position; symbolically we may write:
\begin{equation}
A_{n+1/n+1} (x_1, \ldots, x_n; x) =  
Ad \left[ T_1 (x_1) \ldots T_1(x_n); \Phi_i (x) \right]
\end{equation}
One shows that the perturbative defined object $\Phi_i^{int}$
fulfils the properties like locality and  field equations
in the sense of formal power series. The definition can be regarded as
a direct construction of renormalized composite operators. Epstein
and Glaser showed that the adiabatic limit $g \rightarrow 1$ exists only
in the weak sense of expectation values in massive theories.
The limit possesses all the expected properties of a Green's function
such as causality, Lorentz covariance and the spectral condition. 
\\
$\bullet$ If the specific coupling $T_1$ is fermionic then the
causality condition implies that the corresponding
tempered test function has to be an anticommuting Grassmann variable.
Also an ordering of fermionic couplings in the time-ordered products 
is introduced (for further details see \cite{DHS95,H95}).\\
$\bullet$ The polynomial character of the interactions $T_1$ is forced
by the formalism allowing $g$ to be any element of ${\cal
  S}({\bf{R}}^4)$. If one restricts the choice of the test functions
to the Jaffe class one can also construct theories with
non-polynomial specific couplings (see \cite{EG3}).\\
$\bullet$ Steinmann presented an approach to perturbative quantum field theory 
which is related to the Epstein-Glaser method \cite{steinmann}.
He works  with retarded products and his construction is done
after the adiabatic limit is taken.\\
$\bullet$ A variant of the Epstein-Glaser formalism emerges from the results
of \cite{stora}: The problem of cutting a 
causal distribution into a retarded and advanced  piece 
is equivalent to the problem of continuation of time-ordered
products to coincident points. Actually, the
renormalization scheme of differential renormalization
proposed by Freedman, Johnson and Latorre \cite{freedman} is an
operative way to perform such a continuation in
configuration space (x-space); at least at the one and two-loop level. This 
has been illustrated in
\cite{prange}.
Stora's variant of the EG method allows for an extension of the EG method
to theories on curved space-time. One of the main problems one encounters
in trying to achieve such extension is the absence of translational 
invariance (which plays a crucial role in the original version of
the formalism as presented by Epstein and Glaser (see section 3)) 
in curved space-time. For recent work see \cite{fredenhagen1}.\\ 
$\bullet$ Finally, let us shortly discuss how the problem of
overlapping divergences is automatically solved by the EG
formalism \cite{storapriv}. All $T$-products are defined as operators 
that fulfil Wick's
theorem (\ref{Wick}) and both conditions of causality 
(\ref{caus1}). From this, we can derive a general necessary 
condition for all renormalization schemes which is normally
established at the level of $C$-number valued Green functions and not
at the level of operators. Note that in the EG formalism a Green
function always has a representation as a vacuum expectation value 
of an operator-valued distribution $\< T(V)\>$.

A renormalization of a Green function 
$\< T(V)\>$ has to be such that for all partition of the set of vertices 
$V= X \cup Y, \quad X \neq \emptyset, \quad Y \neq \emptyset$, it
coincides
in the region where $ X \geq Y $ with
\begin{equation}
\label{overlapping}
\< T(V)\> = \< T(X)\> \prod_{x_i \in X, y_i \in Y} D^+(x_i-y_i) \< T(Y)\>
\end{equation}
where  $D^+$ represents a fundamental commutation distribution 
and $\< T(X)\>$ ($\< T(Y)\>$) is a subgraph of order $|X|$ ($|Y|$). In the
EG formalism the latter Green functions are vacuum expectation values
of operator-valued distributions including Wick submonomials (see
(\ref{Wick})).
These are well-defined and fulfil all required conditions  by the 
induction hypothesis. 

Usual renormalization schemes are directly implemented on the level of
$C$-number Green functions via a regularization-substraction scheme.
So it becomes a nontrivial task to show that
all sub-diagrams of a given diagram can be renormalized in a consistent
way, in particular such that the condition (\ref{overlapping}) is fulfilled.
For example, this was shown for Pauli-Villars type regularizations
(see i.e. \cite{EG}).
In the context of the BPHZ method this is achieved by the forest formula 
found by Zimmermann, which solves the recursion formula for the R-operation
as defined in \cite{BS}, an algorithm that 
disentangles all divergences in sub-diagrams.
In the EG formalism it is the inductive operator formalism which 
disentangles the problem of the renormalization of sub-diagrams.

\end{document}